\documentclass[12pt]{JHEP3MaxThesisMod}

\usepackage{amssymb,amsmath,amsfonts}
\usepackage{cite,slashed}
\usepackage{epsfig,subfigure}
\usepackage{epic}

\DeclareMathOperator{\extdm}{d}
\newcommand{\extd}{\extdm \!}

\title{Asymptotic Symmetry Algebras in Non-Anti-de-Sitter Higher-Spin Gauge Theories}
\faculty{Theoretische Physik}
\advisor{{\href{mailto:grumil@hep.itp.tuwien.ac.at}{Daniel Grumiller}}}
\author{{\href{mailto:rieglerm@hep.itp.tuwien.ac.at}{Max Riegler}}\\
	\center{Schwarzspanierstra{\ss}e 15/8/34\\
		1090 Wien}}
\declaration{
 	\begin{flushleft}
    	Ich erkl\"are hiermit, dass ich die eingereichte Masterarbeit selbstst\"andig verfasst habe und keine anderen als die angegebenen Quellen und Hilfsmittel benutzt wurden. Weiters
	versichere ich, dass ich diese Masterarbeit bisher weder im In- noch im Ausland in irgendeiner Form als Pr\"ufungsarbeit vorgelegt habe.\\\vspace{10mm}
  	Wien, \the\year
	\end{flushleft}
  	\begin{flushright}
  	{Max Riegler}
  	\end{flushright}
  	}
\Kurzfassung{Von allen vier fundamentalen Wechselwirkungen ist die Gravitation die einzige Wechselwirkung, von der bis heute keine allgemein akzeptierte quantisierte Theorie existiert.
Ein m\"oglicher L\"osungsansatz zu diesem Problem basiert auf dem holographischen Prinzip. Dieses Prinzip bezeichnet im Wesentlichen die Vermutung, dass eine Theorie der Quantengravitation in $d+1$ Dimensionen durch eine \"aquivalente Beschreibung einer Quantenfeldtheorie (ohne Gravitation) in $d$ Dimensionen formuliert werden kann. In Analogie zu einem Hologramm, bei dem man entweder das abgebildete Objekt in 3 Dimensionen, oder die gespeicherte Intensit\"at und Phase auf dem 2-dimensionalen Schirm als Beschreibung heranziehen kann, so hat man auch im Falle des Holographischen Prinzips zwei gleichwertige, aber dennoch unterschiedliche Beschreibungen der Dynamik eines Systems zur Verf\"ugung.\\
Ein Spezialfall dieses Prinzips, welcher im Kontext der Stringtheorie formuliert wurde, ist die so genannte Anti-de-Sitter/konforme Feldtheorie (AdS/CFT) Korrespondenz, bei der eine Theorie der Quantengraviation mit negativer kosmologischer Konstante durch eine Quantenfeldtheorie, welche unter konformen Transformationen invariant ist, beschrieben werden kann. Da diese Korrespondenz Ergebnisse bei starker Kopplung mit solchen bei schwacher Kopplung verbindet, w\"are dies ein idealer Kandidat, um Quantengravitation auf Skalen, bei denen Quanteneffekte nicht mehr vernachl\"assigt werden k\"onnen, besser zu verstehen. Allerdings gibt es in mehr als 2+1 Dimensionen noch viele konzeptionelle und technische Schwierigkeiten. Daher wird beispielsweise versucht, mit Hilfe von technisch einfacheren Gravitationstheorien in 2+1 Dimensionen zuerst die konzeptionellen Schwierigkeiten zu beseitigen, welche dem Verst\"andnis einer vollst\"andigen Theorie der Quantengravitation im Wege stehen.\\
Viel Aufmerksamkeit in diesem Kontext haben in den vergangenen Jahren auch so genannte H\"ohere-Spin Gravitationstheorien in 2+1 Dimensionen erregt, durch deren Studium man sich ebenfalls ein besseres Verst\"andnis der AdS/CFT Korrespondenz erhofft.\\
In dieser Masterarbeit befassen wir uns mit einer bestimmten H\"ohere-Spin Gravitationstheorie in 2+1 Dimensionen. Wir f\"uhren eine kanonische Analyse durch und stellen konsistente Randbedingen vor, welche von den dynamischen Feldern der Theorie erf\"ullt werden m\"ussen. Weiters bestimmten wir die klassische und quantisierte Symmetriealgebra der daraus resultierenden holographischen Quantenfeldtheorie am Rande der Raumzeit und versuchen sie physikalisch zu interpretieren. 
}       
\abstract{We analyze asymptotic symmetry algebras in (2+1)-dimensional non-AdS higher-spin gravity with a focus on AdS$_2\times\mathbb{R}$ and $\mathbb{H}_2\times\mathbb{R}$. We find a consistent set of boundary conditions for spin-3 gravity in the non-principal embedding and calculate the corresponding asymptotic symmetry algebra in the classical and quantum mechanical case. In addition, we check for unitary representations of the resulting quantum $\mathcal{W}_3^{(2)}$ algebra and give an interpretation of the corresponding CFT.}
\acknowledgements{
           \begin{flushleft}
	At this point I want to thank all the people who supported me in the course of this thesis.\\\vspace{5mm}
	Special thanks go to my supervisor Daniel Grumiller whose support was invaluable. Thank you for your continued guidance since we first met three years ago.\\\vspace{5mm}
	I also want to thank Michael Gary who provided me with a basic mathematica file that I modified to suit my purposes and considerably sped up the process of finding consistent
	boundary conditions for AdS$_2\times\mathbb{R}$ and $\mathbb{H}_2\times\mathbb{R}$.\\\vspace{5mm}
	Furthermore I want to thank Hamid Reza Afshar, Radoslav Rashkov, Sabine Ertl and Stefan Prohazka for enlightening discussions and valuable input.\\\vspace{5mm}
	Very special thanks go to my family and friends and especially to my mother Ulrike and aunt Karin. Without the continuous support of my mother and my aunt it would have been
	impossible for me to finish my studies the way I did. Thank you for everything! I consider myself very lucky to have such an awesome mother and aunt.\\\vspace{10mm}
	Thanks a lot,\\
	Max
	\end{flushleft}}
\begin{document}

\section{Introduction}
In this section we will give a short introduction to the basic concepts underlying this master thesis such as the AdS/CFT correspondence and the motivation to study higher-spin gauge theories.
\subsection{The AdS/CFT Correspondence}
One of the big open questions in physics of the last century is formulating a consistent theory of quantum gravity and in turn maybe also a theory of everything that explains all of the fundamental forces of nature. One possible solution for this problem could be provided by string theory where the fundamental objects are described by one-dimensional objects called \emph{strings} rather than zero-dimensional objects\footnote{For introductional literature on string theory please refer to \cite{Zwiebach:2004,Polchinski:1998rq,Polchinski:1998rr}.}. One conjecture formulated in the framework of string theory and a possible candidate to understand quantum gravity better quantitatively is the so called Anti-de-Sitter/Conformal field theory (AdS/CFT) correspondence. This AdS/CFT correspondence, originally discovered by Maldacena in 1997, is one of the most striking and unexpected discoveries of the last 20 years. Originally formulated as a correspondence between a $N=4$ supersymmetric Yang-Mills theory in four dimensions and a type IIB superstring theory on AdS$_5\times S^5$ \cite{Maldacena:1997re} the correspondence has been much generalized since then and found many applications. The name of the special case of a AdS/CFT correspondence originates from the canonical example according to which the first space is the product of a p+1-dimensional Anti-de-Sitter spacetime\footnote{Anti-de-Sitter spacetimes are maximally symmetric solutions of the Einstein equations with a negative cosmological constant and have constant negative curvature.} and some closed manifold (a sphere for example) and the p-dimensional quantum field theory defined on the boundary is a conformal field theory.\\ 
The generalized conjecture is formulated as an equivalence of a gauge theory (string theory for example) defined on a specific background and a quantum field theory without gravity on the (conformal) boundary of this spacetime. This general principle that the dynamics of a region of spacetime are encoded on the boundary of this region is also called \emph{the holographic principle}. This terminology is indeed adequate because a hologram is completely analogous,  i.e. a three-dimensional image that has been saved on a two dimensional holographic screen but still retaining all information present in three dimensions.\\ 
There is also one prominent physical example that hints to the possibility that the holographic principle is actually present in nature: the entropy of a black hole. Initially black holes were thought of as objects that have zero entropy until Bekenstein \cite{Bekenstein:1981aa} noted that this assumption would violate the second law of thermodynamics. One could for example throw a cup with hot gas and a certain amount of entropy into a black hole and thus decrease the amount of entropy in the universe if the assumption of black holes with zero entropy would be true. Thus, black holes would have to have entropy in order for the second law of thermodynamics to still hold. In fact black holes have more entropy per volume than any other object in the universe. This can be understood by considering a sphere of fixed radius $R$ containing a relativistic gas. The entropy of this gas increases as the energy increases and is only limited by gravitational forces. When the energy exceeds a certain limit the gas collapses to a black hole and thus the resulting black hole has to contain at least the same amount of entropy as the gas before the collapse. Bekenstein used this argument to conjecture an upper bound of the entropy of a black hole which is proportional to the area of the black hole. This conjecture was later confirmed by Hawking \cite{Hawking:1975sw}. Since in statistical physics entropy is proportional to the logarithm of the number of possible microstates, the Bekenstein-Hawking entropy suggests that the logarithm of the number of microstates of a black hole is proportional to its area rather than its volume. This is a statement that strongly hints at the validity of the holographic principle.\\
Now back to the main reason why this AdS/CFT correspondence is a candidate do deepen our understanding of quantum gravity. This duality is a strong/weak duality. This means that the coupling constants of the bulk and boundary theories are related in such a way that if one tuned the coupling of the bulk theory such that the theory is strongly coupled, then the dual boundary theory would be weakly coupled and vice versa. This can be seen for example by considering the perturbative expansion of the partition function of a large $N$ gauge theory and the loop expansion in string theory. The perturbative expansion of a large $N$ gauge theory in $\frac{1}{N}$ and $g_{YM}^2N$ is given by \cite{deBoer:2003aa}
	\begin{equation}\label{Intro:PartitionSYM}
		Z=\sum_{g\geq0}N^{2-2g}f_g(\lambda),
	\end{equation}
where $g_{YM}$ is the coupling constant of the (Yang-Mills) gauge theory, $\lambda=g_{YM}^2N$ is the so called 't Hooft coupling and $f_g(\lambda)$ are arbitrary functions of  $\lambda$. This expansion looks very much like the loop expansion in string theory given by
	\begin{equation}
		Z=\sum_{g\geq0}g_s^{2g-2}Z_g,
	\end{equation}
where $g_s$ is the string coupling which is identified with $\frac{1}{N}$ in large $N$ dualities. Considering the supergravity approximation of string theory on AdS$_5\times S^5$ this theory has three important parameters:
	\begin{itemize}
		\item The string coupling constant $g_s$ which measures the string interaction strength relevant for splitting and joining of strings.
		\item The string length $l_s$ measuring the size of the fluctuations of the string worldsheet.
		\item The curvature radius $\ell$ of AdS$_5$ and $S^5$.
	\end{itemize}
Four dimensional $\mathcal{N}=4$ super Yang-Mills (SYM) Theory\footnote{$\mathcal{N}$ is the number of supersymmetry generators.} with $U(N)$ gauge group has the following two parameters:
	\begin{itemize}
		\item The rank $N$ of the gauge group.
		\item The coupling constant $g_{YM}$ determining the strength of interactions in SYM theories.
	\end{itemize}
These parameters are related as follows
	\begin{equation}
		g_s=g_{YM}^2,\quad\left(\frac{\ell}{l_s}\right)^4=4\pi g_{YM}^2N=4\pi\lambda.
	\end{equation}
Looking at \eqref{Intro:PartitionSYM} we see that the SYM theory perturbative expansion is only valid for small $g_{YM}$ and small $\lambda$. Since the supergravity approximation of string theory is only valid for $\frac{\ell}{l_s}\gg1$, which is equivalent to $\lambda\gg1$, we see that we have indeed a strong/weak duality. If 
	\begin{itemize}
		\item $\lambda\ll1$ then the SYM theory is weakly coupled but the dual string theory is strongly coupled,
		\item $\lambda\gg1$ then the SYM theory is strongly coupled and the dual string theory is weakly coupled.
	\end{itemize}
Thus if $\lambda\gg1$ for example then it would be possible to compute certain observables on the string theory side perturbatively and then translate these results to the SYM side and thus gain results for observables in the strongly coupled regime of this theory. In order to establish some kind of dictionary that would help in this translation process it is thus helpful to find dual theories that are completely solvable. Excellent candidates on the field theory side are thus 2-dimensional CFTs since these theories are exactly solvable in generic regimes of parameter space. This is a direct consequence of the fact that the algebra of infinitesimal conformal transformations is infinite dimensional in two dimensions\footnote{As an introduction to CFTs please refer to \cite{diFrancesco,Blumenhagen,Schottenloher,Ginsparg:1988ui}.}. Thus CFTs in two dimensions are ideal models to address fundamental questions about quantum gravity that are usually very hard to answer in higher dimensional field theories. As a consequence we have to consider gravity theories in three dimensions as appropriate dual theories to two dimensional field theories.
\subsection{Einstein-Hilbert Gravity as a Chern-Simons Theory}\label{Intro:EHGAACST}
In this section we try to give a short introduction as to how it is possible to reformulate Einstein-Hilbert gravity in 2+1 dimensions as a Chern-Simons gauge theory.\\
2+1 dimensional pure gravity without matter fields is described by the Einstein-Hilbert action
	\begin{equation}
		I_{EH}=\frac{1}{16\pi G_N}\int_{\mathcal{M}}\extd^3x\sqrt{-g}\left(R+\frac{2}{\ell^2}\right),
	\end{equation}
with $G_N$ being Newton's constant, $R$ the Ricci scalar and $\frac{1}{\ell^2}=-\Lambda$ is the cosmological constant where $\ell$ denotes the AdS radius. One of the main problems one encounters when trying to quantize gravity in dimensions $d>3$ is the problem of non-renormalizability. The infinities that usually occur when one expands gravity in terms of Feynman diagrams cannot be absorbed by a renormalization of the gravitational coupling constant \cite{Maldacena:1999fi}. This means that one would have to determine an infinite amount of parameters in order to fully describe quantum gravity. Following this argument it was also believed that gravity in $D=2+1$ is non-renormalizable. However, the theory in 2+1 dimensions is trivial in the bulk on the classical level in the sense that there are no propagating local degrees of freedom, i.e. gravitons. Since quantized theories that are trivial on the classical level are usually renormalizable there was a high probability that gravity in 2+1 dimensions is actually renormalizable. And indeed it was then shown by Witten in \cite{Witten:1988hc} that in 2+1 dimensions renormalization is indeed possible. The main reason for this is that in 2+1 dimensions the Riemann tensor $R_{abcd}$ can be expressed in terms of the Ricci tensor $R_{ab}$, the Ricci scalar $R$ and the metric $g_{ab}$ as
	\begin{equation}
		R_{abcd}=g_{ac}R_{bd}+g_{bd}R_{ac}-g_{ad}R_{bc}-g_{bc}R_{ad}-\frac{1}{2}R(g_{ac}g_{bd}-g_{ad}g_{bc}).
	\end{equation}
Since the equations of motion of pure Einstein-Hilbert gravity given by
	\begin{equation}
		R_{ab}-\frac{1}{2}g_{ab}\left(R+\frac{2}{\ell^2}\right)=0
	\end{equation}
tell us that the Ricci tensor is equal to some scalar times the metric one can express all curvature invariants $R_{abcs}$, $R_{ab}$ and $R$ in terms of the metric $g_{ab}$. This in turn implies that all possible counterterms\footnote{The lowest order pure curvature corrections to the Einstein-Hilbert action look for example like $R\sqrt{|g|}$, $R_{ab}R^{ab}\sqrt{|g|}$, $R_{abcd}R^{abcd}\sqrt{|g|}$.}, which should cancel the infinite diagrams can be written as a multiple of $\int\extd^3x\sqrt{|g|}$, which is equivalent to an on shell renormalization of the cosmological constant. If there are counterterms present that vanish on shell then these terms can be absorbed by a redefinition of the metric \cite{'tHooft:1974bx} $g_{ab}\rightarrow g+\alpha R_{ab}+\ldots$, where $\alpha$ is some constant and the ellipsis denotes higher order terms of curvature invariants. Thus in 2+1 dimensions all divergencies in perturbation theory can be removed by a redefinition of the metric and the cosmological constant \cite{Witten:2007kt}.\\ 
In order to see that pure Einstein gravity and a Chern-Simons theory in three dimensions are equivalent up to boundary terms, it is convenient to formulate 2+1 dimensional general relativity in terms of a local orthonormal basis for the (co)tangent space called dreibein $e$, which can be interpreted as a local inertial frame and a spin connection $\omega$. The dreibein and spin connection in terms of a cotangent basis are given by
	\begin{equation}
		e^a=e^a{}_\mu dx^\mu,\quad\omega^{ab}=\omega^{ab}{}_\mu dx^\mu.
	\end{equation}
Latin letters $a,b,\ldots$ denote local Lorentz indices, greek letters $\mu,\nu,\ldots$ denote spacetime indices and $\omega^{ab}{}_\mu=-\omega^{ba}{}_\mu$.
The spacetime metric and the dreibein are related as
	\begin{equation}\label{Intro:MetricDreibeinRel}
		g_{\mu\nu}=e^a{}_\mu e^b{}_\nu\eta_{ab},
	\end{equation}
where $\eta_{ab}$ denotes the Minkowski metric with signature $(-,+,+)$. A very convenient feature of 2+1 dimensions is that one can "dualize" the spin connection in the following way 
	\begin{equation}\label{Intro:DualizedSpinConn}
		\omega^a=\frac{1}{2}\epsilon^{abc}\omega_{bc},
	\end{equation}
where $\epsilon^{abc}$ is the Levi-Civita symbol and we omitted the spacetime index $\mu$ for the sake of brevity. The curvature 2-form in terms of the dualized spin connection is then given by
	\begin{equation}
		R^a=d\omega^a+\epsilon^a{}_{bc}\omega^b\wedge\omega^c.
	\end{equation}
One can now regard the dreibein $e$ and the spin connection $\omega$ as the new dynamical variables of the theory and reformulate the Einstein-Hilbert action in terms of these new variables. 
	\begin{equation}\label{Intro:IEHP}
		I_{EHP}=\frac{1}{16\pi G_N}\int_{\mathcal{M}}R^a\wedge e_a +\frac{2}{3\ell^2}\epsilon_{abc}e^a\wedge e^b\wedge e^c,
	\end{equation}
The action \eqref{Intro:IEHP} already looks very familiar in comparison to the Chern-Simons action
	\begin{equation}
		I_{CS}[A]=\frac{k}{4\pi}\int_\mathcal{M}\textnormal{Tr}(A\wedge\extd A+\frac{2}{3}A\wedge A\wedge A),
	\end{equation}
where $A$ is a Lie algebra valued one form, which in the case of pure Einstein-Hilbert gravity will be $\mathfrak{sl}(2)$. Combining the dreibein and the dualized spin connection into the following connection one forms
	\begin{align}
		A^a=\omega^a+\frac{1}{\ell}e^a\\
		\bar{A}^a=\omega^a-\frac{1}{\ell}e^a,
	\end{align}
Witten showed in \cite{Witten:1988hc} that the combination
	\begin{equation}
		I=I_{CS}[A]-I_{CS}[\bar{A}]
	\end{equation}
is equivalent to the Einstein-Hilbert action up to boundary terms. Let $L_a$, $a=-1,0,1$ denote the $\mathfrak{sl}(2)$ generators then the normalization \cite{Campoleoni:2011hg}
	\begin{equation}
		\textnormal{Tr}(L_aL_b)=\frac{1}{2}\eta_{ab}
	\end{equation}
leads to the following identification 
	\begin{equation}
		k=\frac{\ell}{4G_N}.
	\end{equation}
In a similar fashion one can construct a first order formulation of the dynamics of free massless bosonic symmetric fields with spin $s\geq2$ \cite{Vasiliev:1980as,Lopatin:1987hz}. There is, however, one notable subtlety in comparison to the $\mathfrak{sl}(2)$ construction. Dualizing the spin connection as in \eqref{Intro:DualizedSpinConn} only works for gauge groups that have dimension $3$, like $\mathfrak{sl}(2)$. Thus in higher-spin gravity one starts with $A^a$ and $\bar{A}^a$ and defines the corresponding spin connection and zuvielbein\footnote{Since in general for higher-spin theories the Lie algebras considered have more than three generators and thus do not have to match the number of spacetime indices, $e^a{}_\mu$ is called zuvielbein rather than dreibein, as it was called in the $\mathfrak{sl}(2)$ case.} as
	\begin{align}
		e^a=\frac{\ell}{2}(A^a-\bar{A}^a)\\
		\omega^a=\frac{1}{2}(A^a-\bar{A}^a).
	\end{align}
The spacetime metric is then defined as
	\begin{equation}\label{Intro:HigherSpinMetric}
		g_{\mu\nu}:=\left(\#\right)\textnormal{Tr}\left(A_\mu-\bar{A}_\mu\right)\left(A_\nu-\bar{A}_\nu\right),
	\end{equation}
where $\left(\#\right)$ is some convenient factor of normalization. This definition is one possibility in linking the gauge potentials $A$ and $\bar{A}$ and the spacetime metric $g_{\mu\nu}$. This is a viable choice since the metric defined in this way is manifestly gauge invariant and in the case of $\mathfrak{sl}(2)$ coincides with \eqref{Intro:MetricDreibeinRel}. It is, however, also possible that one could add further terms to \eqref{Intro:HigherSpinMetric} and thus a unique definition of the spacetime metric in terms of the gauge potentials $A$ and $\bar{A}$ is still an open topic as of yet. Thus taking these subleties into account one can construct bosonic massless higher-spin gauge theories via $G\times G$ Chern-Simons theories \cite{Blencowe:1988gj}, where $G$ denotes the gauge group of the theory.\\
As already mentioned, gravity theories as well as Chern-Simons theories are trivial in the sense that they have no propagating local degrees of freedom. This does, however, not mean that these theories are trivial. In fact they are quite nontrivial as soon as one introduces a boundary. In general there is even a symmetry enhancement of the bulk symmetries occurring at the boundary. The by now famous example of a $SL(2)\times SL(2)$ bulk isometry algebra that is enhanced to two copies of a Virasoro algebra with a central charge $c=6k$ at the boundary has first been studied by Brown and Henneaux \cite{Brown:1986nw}. By adding further massless higher-spin excitations the asymptotic symmetries are even further enhanced to non linear algebras called $\mathcal{W}$-algebras\footnote{For an introduction to $\mathcal{W}$-algebras please refer to \cite{Bouwknegt:1992wg}.}.
\subsection{Higher-Spin Gravity}
In this section we will motivate why it is interesting to study higher-spin gravity in 2+1 dimensions.\\
Higher-spin excitations appear quite naturally in (super)string theories. In addition to massless modes of lower spin $s\leq2$ there is an infinite tower of massive modes of arbitrary high spin with their mass squared proportional to the string tension $T$ and spin $s$. Since the string tension is inverse proportional to the string length squared, these higher-spin modes are very heavy and thus unobservable at low energies. Nevertheless these higher-spin excitations are necessary for the consistency of a string theory describing all the fundamental interactions. In general, quantum field theories containing massive particles with spin $s\geq1$ are non-renormalizable unless the mass was acquired through some kind of spontaneous symmetry breaking. Thus it may be possible that string theory is just a broken phase of some other gauge theory with additional higher-spin symmetry and corresponding massless higher-spin gauge fields \cite{Sorokin:2005aa}. Since the mass squared of the higher-spin excitations is proportional to the string tension, these modes become massless in the limit $T\rightarrow0$. Thus, in this limit one should observe a symmetry enhancement of string theory by higher-spin symmetry, and one can regard string tension generation as a mechanism of the spontaneous symmetry breaking of the higher-spin symmetry. If this conjecture was true, then this could be very useful in understanding string theory and in particular the AdS/CFT correspondence. However, it is not easy to build such theories with higher-spin gauge symmetries in flat space. There is in particular one theorem by Coleman and Mandula which has been generalized to arbitrary dimensions by Pelc and Horwitz \cite{Pelc:1997aa}, which states that symmetries of the S-matrix in a non-trivial (i.e. interacting) field theory in a
flat space can only have sufficiently low spins. This theorem can, however, be circumvented if one is considering AdS spacetimes \cite{Vasiliev:1999aa}. And indeed for AdS there exist interacting higher-spin theories of massless particles \cite{Fradkin:1987aa,Fradkin:1987ab,Vasiliev:1990aa,Vasiliev:2003aa}. There exists also a general statement that the cosmological constant $\Lambda$ in dimensions $D>2+1$ should be non-zero in the phase of the unbroken higher-spin symmetry \cite{Vasiliev:1990en,Vasiliev:2003aa}, thus considering AdS spacetimes is not simply a trick to circumvent the Coleman-Mandula theorem as it might seem at first. One could for example start with a theory of massless higher-spin fields and a cosmological constant $\Lambda\neq0$. After spontaneous symmetry breaking via some mechanism (dimensional compactification for example) one could indeed end up in principle with a theory where $m\neq0$ for higher-spin fields and $\Lambda=0$ (or $\Lambda$ very small) \cite{Vasiliev:1999ba}. The modification of the cosmological constant could then be due to some fields that acquire a non-zero vacuum expectation value via the spontaneous symmetry breaking and thus modify the vacuum energy\footnote{This is due to the fact that a cosmological constant has the same effect as an intrinsic energy density of the vacuum.}. Thus, it would be possible in principle to start with a massless higher-spin theory and $\Lambda\neq0$, and after spontaneous symmetry breaking one could end up with a string theory containing massive higher-spin fields and a very small cosmological constant.\\
In general, higher-spin gauge theories contain an infinite set of spins $0\leq s\leq\infty$. Thus it is generically not possible to just consider particles up to spin $n$ in a higher-spin gauge theory. The only known exception to this is the case of $2+1$-dimensions, where it is possible to truncate this otherwise infinite tower of higher-spin fields at arbitrary spin $n$ so that all fields have spin $s\leq n$ \cite{Aragone:1984aa}, which is another reason why it is interesting to work in 2+1 dimensions.
\subsection{Non-AdS Holography}
There are many applications that require a generalization of the AdS/CFT correspondence to a gauge/gravity duality that does not involve spacetimes asymptoting to AdS, or asymptoting to AdS in a weaker way as compared to Brown-Henneaux boundary conditions \cite{Brown:1986nw,Henneaux:2010aa}. Some examples are given by
	\begin{itemize}
		\item  \emph{null warped AdS} spacetimes, which arise in proposed holographic duals of non-relativistic CFTs describing cold atoms \cite{Son:2008ye,Balasubramanian:2008dm}
		\item \emph{Schr\"odinger} spacetimes,  which generalize null warped AdS by introducing an arbitrary scaling exponent \cite{Adams:2008wt}
		\item \emph{Lifshitz} spacetimes, which arise in gravity duals of Lifshitz-like fixed points \cite{Kachru:2008yh} and also have a scaling exponent parametrizing spacetime anisotropy
		\item \emph{AdS/log CFT} correspondence \cite{Grumiller:2008qz,Ertl:2009ch}, which requires a relaxation of the Brown-Henneaux boundary conditions \cite{Grumiller:2008es,Henneaux:2009aa,Maloney:2010aa}
		\item\emph{Flat space holography}, which requires the spacetime in the bulk to be asymptotically flat \cite{Susskind:1998vk,Arcioni:2003xx,Arcioni:2003td}.
	\end{itemize}
A priori it is not clear that higher-spin gravity can accommodate such non-AdS backgrounds. There is, however, one example of a theory that is very similar to higher-spin gravity in 2+1 dimensions, namely conformal Chern-Simons gravity \cite{Deser:1982vy,Deser:1982wh,Horne:1988jf}. Conformal Chern-Simons gravity has
	\begin{itemize}
		\item no local physical degrees of freedom,
		\item a Chern-Simons formulations with a gauge group bigger than $SL(2)\times SL(2)$,
		\item gauge symmetries that relate non-diffeomorphic metrics to each other,
		\item and the asymptotic symmetry group can be larger than two copies of the Virasoro algebra \cite{Afshar:2011yh,Afshar:2011qw}.
	\end{itemize}
The axisymmetric stationary solutions of conformal Chern-Simons gravity include AdS$_3$ as well as AdS$_2\times\mathbb{R}$, which means that at least for conformal Chern-Simons gravity non-AdS backgrounds exist. And indeed it has been shown in \cite{Gary:2012aa} that higher-spin gravity with an appropriate variational principle is indeed capable of generating spacetimes that asymptote to AdS (with weaker boundary conditions than Brown-Henneaux), AdS$_2\times\mathbb{R}$, Schr\"odinger, Lifshitz and warped AdS spacetimes.

\clearpage

\section{Basics of Chern-Simons Theories}
In this section we will give a short introduction to the basics of Chern-Simons theories and explain the variational principle we will be using in order to accommodate asymptotic backgrounds beyond AdS. Since there already exist excellent books explaining the basics of constrained hamiltonian systems and canonical analysis, we will not go much into detail regarding these topics. We refer the interested reader to \cite{Henneaux:1992,Blagojevic:2002aa} for example.
\subsection{Chern-Simons Action}
As reviewed in section \ref{Intro:EHGAACST} Einstein gravity with a negative cosmological constant in three dimensions can be reformulated as the difference of two Chern-Simons actions given by
	\begin{equation}
		I=I_{CS}[A]-I_{CS}[\bar{A}]
	\end{equation}
with
	\begin{equation}\label{Intro:ICS}
		I_{CS}[A]=\frac{k}{4\pi}\int_{\mathcal{M}}\textnormal{Tr}(A\wedge\extd A+
		\frac{2}{3}A\wedge A\wedge A)+B[A].
	\end{equation}
$I_{CS}[\bar{A}]$ is given by just replacing $A\rightarrow\bar{A}$ in \eqref{Intro:ICS}. Hence we will focus on the canonical analysis of the $I_{CS}[A]$ term. The canonical analysis for $I_{CS}[\bar{A}]$ can then be obtained in complete analogy to the one performed with $I_{CS}[A]$ just by replacing $k\rightarrow-k$ and $A\rightarrow\bar{A}$ in all relevant formulas. In addition we will also set the AdS radius $\ell$ to 1.\\
The action given by \eqref{Intro:ICS} is defined on a Manifold $\mathcal{M}$ with the topology $\mathcal{M}=\mathbb{R}\times\Sigma$ and coordinates $x^\mu=(t,\rho,\varphi)$, $\mu=0,1,2$. In addition, we assume that $\Sigma$ has the topology of a disk and is parameterized by $\varphi$ and $\rho$, where $\rho=$ const. corresponds to the boundary. $B[A]$ is a boundary term defined on 
$\partial\mathcal{M}=\partial\Sigma\times\mathbb{R}$ to ensure a well defined variational principle and gauge invariance of the action if one wants to consider spacetimes that do not asymptote to AdS. Without this boundary term the resulting restrictions on the connection $A$ that would ensure a well defined variational principle and gauge invariance of the action would only allow the resulting spacetimes to asymptote to AdS$_3$. The $A$'s are Lie algebra valued 1-forms that can be written as
	\begin{equation}
		A=A^a{}_\mu\extd x^\mu T_a,
	\end{equation}
with $T_a$ being a basis of the Lie algebra $\mathfrak{g}$ one is considering
. If one chooses such a basis then $\textnormal{Tr}(T_aT_b)$ can be interpreted as a non-degenerate bilinear form on the Lie algebra. In components one can write \eqref{Intro:ICS} as
	\begin{equation}\label{Intro:CoordAction}
		I_{CS}[A]=\frac{k}{4\pi}\int_{\mathcal{M}}\extd^3x\epsilon^{\mu\nu\lambda}g_{ab}\left(
		A^a{}_\mu\partial_\nu A^b{}_\lambda+\frac{1}{3}f^a{}_{cd}A^c {}_\mu A^d{}_\nu A^b{}_\lambda\right)
		+B[A],
	\end{equation}
where $g_{ab}=\textnormal{Tr}(T_aT_b)$, $\epsilon^{t\rho\varphi}=1$ and $f^a{}_{bc}$ are the structure constants of the Lie algebra given by
	\begin{equation}
		\left[T_a,T_b\right]=f^c{}_{ab}T_c.
	\end{equation}
Lie algebra indices $(a,b,\ldots)$ are raised and lowered with $g_{ab}$ and spacetime indices $(\mu,\nu,\ldots)$ with the background metric $g_{\mu\nu}$ of the spacetime considered.\\ 
\subsection{Variational Principle and Equations of Motion}
In order to obtain the equations of motion one has to vary \eqref{Intro:ICS}. This yields
	\begin{equation}
		\delta I_{CS}[A]=\frac{k}{2\pi}\int_{\mathcal{M}}\textnormal{Tr}(\delta A\wedge F)+\frac{k}{4\pi}\int_{\partial\mathcal{M}}\textnormal{Tr}(\delta A\wedge A)+
				     \delta B[A],
	\end{equation}
with $F=dA+A\wedge A$. One could consider for example the following boundary term
	\begin{equation}
		B[A]=\frac{k}{4\pi}\int_{\partial\mathcal{M}}\extd^2x\,\textnormal{Tr}(A_\varphi A_t).
	\end{equation}
In order to have a well defined variational principle we have to require $\delta I_{CS}[A]=0$ which specifies the equations of motion as
	\begin{equation}
		F=0.
	\end{equation} 
In addition, we have to restrict the boundary conditions such that the total boundary term vanishes, i.e.
	\begin{equation}
		\delta I_{CS}[A]\Bigr|_{\textnormal{on-shell}}=
		\frac{k}{2\pi}\int_{\partial\mathcal{M}}\extd^2x\,g_{ab}A^a{}_\varphi\delta A^b{}_t=0.
	\end{equation}
This can be achieved by demanding either
	\begin{equation}
		A_\varphi\Bigr|_{\partial\mathcal{M}}=0\quad\textnormal{or}\quad\delta A_t\Bigr|_{\partial\mathcal{M}}=0.
	\end{equation}
Since $A_\varphi\Bigr|_{\partial\mathcal{M}}=0$ is a slightly stronger boundary condition on the connection than $\delta A_t\Bigr|_{\partial\mathcal{M}}=0$, we will use the latter one because we do not want to put too many restrictions on the connection.\\
\subsection{Gauge Invariance of the Chern-Simons Action}
Another important consistency condition is gauge invariance of the action. Since the connection has to satisfy $\delta A_t\Bigr|_{\partial\mathcal{M}}=0$, the form of the allowed gauge transformations will also necessarily be restricted. Writing finite gauge transformations as
	\begin{equation}\label{Intro:FiniteGaugeTrafo}
		A\rightarrow g^{-1}(\tilde{A}+d)g,
	\end{equation}
with $g\in G$ where $G$ is the gauge group one is considering, we can calculate the change of the action \eqref{Intro:ICS} under \eqref{Intro:FiniteGaugeTrafo}. This leads to
	\begin{equation}
		I_{CS}[A]=I_{CS}[\tilde{A}]+\delta I_{CS}[\tilde{A}]+\delta B[\tilde{A}],
	\end{equation}
where
	\begin{equation}\label{Intro:GaugeI}
		\delta I_{CS}[\tilde{A}]=
		-\frac{k}{12\pi}\int_{\mathcal{M}}\textnormal{Tr}(g^{-1}dg\wedge g^{-1}dg\wedge g^{-1}dg)
		-\frac{k}{4\pi}\int_{\partial\mathcal{M}}\textnormal{Tr}(dgg^{-1}\wedge\tilde{A})
	\end{equation}
and
	\begin{equation}\label{Intro:GaugeB}
		\delta B[\tilde{A}]=-\frac{k}{4\pi}\int_{\partial\mathcal{M}}\extd^2x
		\textnormal{Tr}(\partial_\varphi g\partial_tg^{-1}-\tilde{A}_\varphi\partial_tgg^{-1}-\tilde{A}_t\partial_\varphi gg^{-1}).
	\end{equation}
Hence the Chern-Simons action is gauge invariant if either
	\begin{itemize}
		\item $g\rightarrow1$ sufficiently fast when approaching the boundary $\partial\mathcal{M}$; 
		or
		\item the gauge transformations are certain infinitesimal gauge transformations as specified below.
	\end{itemize}
Infinitesimal gauge transformations connected to the identity are given by
	\begin{equation}
		g\simeq1+\lambda^aT_a.
	\end{equation}
This leads to
	\begin{align}
		\delta I_{CS}[\tilde{A}]+\delta B[\tilde{A}]=&-\frac{k}{4\pi}\int_{\partial\mathcal{M}}\extd^2x\,
		g_{ab}\left(\epsilon^{\rho ij}\partial_i\lambda^a\tilde{A}^b{}_j-\tilde{A}^a{}_\varphi\partial_t\lambda^b-\tilde{A}^a{}_t\partial_\varphi\lambda^b\right)\nonumber\\
		=&\frac{k}{2\pi}\int_{\partial\mathcal{M}}\extd^2x\,g_{ab}\tilde{A}^b{}_\varphi\partial_t\lambda^a=0.
	\end{align}
Since we do not want to impose additional constraints on the connection, one can conclude that \eqref{Intro:ICS} is gauge invariant for infinitesimal gauge transformations satisfying at the boundary
	\begin{equation}
		\partial_t\lambda^a\Bigr|_{\partial\mathcal{M}}=0.
	\end{equation}
\subsection{Canonical Analysis of Chern-Simons Theories}
In order to proceed with the canonical analysis it is convenient to use a $2+1$ decomposition of the action \eqref{Intro:CoordAction} \cite{Banados:1994tn}. The $2+1$ decomposition of \eqref{Intro:CoordAction} is given by
	\begin{equation}\label{Intro:2+1}
		I_{CS}[A]=\frac{k}{4\pi}\int_{\mathbb{R}}\extd t\int_{\Sigma}\extd^2x\epsilon^{ij}g_{ab}\left(\dot{A}^a{}_iA^b{}_j+A^a{}_0F^b{}_{ij}+\partial_j
			\left(A^a{}_iA^b{}_0\right)\right)+B[A],
	\end{equation}
with $F^a{}_{ij}=\partial_iA^a{}_j-\partial_jA^a{}_i+f^a{}_{bc}A^b{}_iA^c{}_j$ and $\epsilon^{ij}=\epsilon^{tij}$. Since the EOM require $F^a {}_{ij}=0$, the form of \eqref{Intro:2+1} already specifies $A^a_0$ as a Lagrange multiplier and $A^a_i$ as the dynamical fields. The Lagrangian density $\mathcal{L}$ is then given by
	\begin{equation}
		\mathcal{L}=\frac{k}{4\pi}\epsilon^{ij}g_{ab}\left(\dot{A}^a{}_iA^b{}_j+A^a{}_0F^b{}_{ij}+\partial_j\left(A^a{}_iA^b{}_0\right)\right).
	\end{equation}
Calculating the canonical momenta $\pi_a{}^\mu\equiv\frac{\partial\mathcal{L}}{\partial\dot{A}^a_\mu}$ corresponding to the canonical variables $A^a_\mu$ one finds the following primary constraints
	\begin{equation}
		\phi_a{}^0:=\pi_a{}^0\approx0\quad\phi_a{}^i:=\pi_a{}^i-\frac{k}{4\pi}\epsilon^{ij}g_{ab}A^b{}_j\approx0.
	\end{equation}
The Poisson brackets of the canonical variables are given by
	\begin{equation}\label{Intro:CanComm}
		\{A^a{}_\mu(\textbf{x}),\pi_b{}^\nu(\textbf{y})\}=\delta^a{}_b\delta_\mu{}^\nu\delta^2(\textbf{x}-\textbf{y}).
	\end{equation}
The next step is to calculate the canonical Hamiltonian density via the following Legendre transformation 
	\begin{equation}
		\mathcal{H}=\pi_a{}^\mu\dot{A}^a{}_\mu-\mathcal{L}=-\frac{k}{4\pi}\epsilon^{ij}g_{ab}\left(A^a{}_0F^b{}_{ij}+\partial_j\left(A^a{}_iA^b{}_0\right)\right).
	\end{equation}
Since we are dealing with a constrained Hamiltonian system, we have to work with the total Hamiltonian given by
	\begin{equation}
		\mathcal{H}_T=\mathcal{H}+u^a{}_\mu\phi_a{}^\mu,
	\end{equation} 
where $u^a{}_\mu$ are some arbitrary multipliers. Since the primary constraints should be conserved after a time evolution, we require
	\begin{equation}
		\dot{\phi}_a{}^\mu=\{\phi_a{}^\mu,\mathcal{H}_T\}\approx0,
	\end{equation}
which leads to the following secondary constraints
	\begin{align}
		\mathcal{K}_a\equiv-\frac{k}{4\pi}\epsilon^{ij}g_{ab}F^b{}_{ij}&\approx0\\
		D_iA^a{}_0-u^a{}_i&\approx0,\label{Intro:Multiplier}
	\end{align}
where $D_iX^a=\partial_iX^a+f^a{}_{bc}A^b{}_iX^c$ is the covariant derivative. One can now use the Hamilton equations of motion, which are given by
	\begin{equation}
		\dot{A}^a{}_i=\frac{\partial\mathcal{H}_T}{\partial\pi_a{}^i}=u^a{}_i
	\end{equation}
to determine the Lagrange multipliers $u^a{}_i$ and rewrite \eqref{Intro:Multiplier}. This yields the following weak equality
	\begin{equation}
		D_iA^a{}_0-u^a{}_i = D_iA^a{}_0-\partial_0A^a{}_i = F^a{}_{i0}\approx0.
	\end{equation}
The total Hamiltonian can now be written in the following form
	\begin{equation}
		\mathcal{H}_T=A^a{}_0\bar{\mathcal{K}}_a+u^a{}_0\phi_a{}^0+\partial_i(A^a{}_0\pi_a{}^i),
	\end{equation}
with
	\begin{equation}
		\bar{\mathcal{K}}_a=\mathcal{K}_a-D_i\phi_a{}^i.
	\end{equation}
One can use the canonical commutation relations \eqref{Intro:CanComm} to determine the following Poisson brackets which will be necessary to determine the Poisson algebra of the constraints
	\begin{subequations}\label{Intro:ConstraintBrackets}
		\begin{align}
			\{\phi_a{}^0(\textbf{x}),A^b{}_0(\textbf{y})\}&=-\delta_a{}^b\delta^2(\textbf{x}-\textbf{y}),\\
			\{\phi_a{}^i(\textbf{x}),A^b{}_j(\textbf{y})\}&=-\delta_a{}^b\delta^i{}_j\delta^2(\textbf{x}-\textbf{y}),\\
			\{\phi_a{}^i(\textbf{x}),\pi_b{}^j(\textbf{y})\}&=-\frac{k}{4\pi}\epsilon^{ij}g_{ab}\delta^2(\textbf{x}-\textbf{y}),\\
			\{\phi_a{}^i(\textbf{x}),\pi_b{}^j(\textbf{y})\}&=-\frac{k}{2\pi}\epsilon^{ij}g_{ab}\delta^2(\textbf{x}-\textbf{y}),\\
			\{A^a{}_i(\textbf{x}),D_j\phi_b{}^j(\textbf{y})\}&=[\delta^a{}_b\partial_i+f^a{}_{bc}A^c{}_i(\textbf{y})]\delta^2(\textbf{x}-\textbf{y}),\\
			\{\pi_a{}^i(\textbf{x}),D_j\phi_b{}^j(\textbf{y})\}&=-\frac{k}{4\pi}\epsilon^{ij}[g_{ab}\partial_j+f_{abc}A^c{}_j(\textbf{y})]\delta^2(\textbf{x}-\textbf{y})
										+f_{ab}{}^c\phi_c{}^i(\textbf{y})\delta^2(\textbf{x}-\textbf{y}),\\
			\{\phi_a{}^i(\textbf{x}),D_j\phi_b{}^j(\textbf{y})\}&=-\frac{k}{2\pi}\epsilon^{ij}[g_{ab}\partial_j+f_{abc}A^c{}_j(\textbf{y})]\delta^2(\textbf{x}-\textbf{y})
										+f_{ab}{}^c\phi_c{}^i(\textbf{y})\delta^2(\textbf{x}-\textbf{y}),\\
			\{\pi_a{}^i(\textbf{x}),\mathcal{K}_b(\textbf{y})\}&=-\frac{k}{2\pi}\epsilon^{ij}[g_{ab}\partial_j+f_{abc}A^c{}_j(\textbf{y})]\delta^2(\textbf{x}-\textbf{y}),\\
			\{\phi_a{}^i(\textbf{x}),\mathcal{K}_b(\textbf{y})\}&=-\frac{k}{2\pi}\epsilon^{ij}[g_{ab}\partial_j+f_{abc}A^c{}_j(\textbf{y})]\delta^2(\textbf{x}-\textbf{y}),\\
			\{D_i\phi_a{}^i(\textbf{x}),\mathcal{K}_b(\textbf{y})\}&=-\frac{k}{2\pi}\epsilon^{ij}f_{abc}D_iA^c{}_j\delta^2(\textbf{x}-\textbf{y}),\\
			\{\phi_a{}^i(\textbf{x}),\bar{\mathcal{K}}_b(\textbf{y})\}&=-f_{ab}{}^c\phi_c{}^i\delta^2(\textbf{x}-\textbf{y}),\\
			\{D_i\phi_a{}^i(\textbf{x}),D_j\phi_b{}^j(\textbf{y})\}&=-\frac{k}{2\pi}\epsilon^{ij}f_{abc}D_iA^c{}_j\delta^2(\textbf{x}-\textbf{y})-
											f_{ab}{}^cD_i\phi_c{}^i\delta^2(\textbf{x}-\textbf{y}),
		\end{align}
	\end{subequations}
where $\partial_i$ denotes $\frac{\partial}{\partial y^i}$. Using these relations one finds the following algebra of constraints
	\begin{subequations}
		\begin{align}
			\{\phi_a{}^i(\textbf{x}),\phi_b{}^j(\textbf{y})\}&=-\frac{k}{2\pi}\epsilon^{ij}g_{ab}\delta^2(\textbf{x}-\textbf{y}),\\
			\{\phi_a{}^i(\textbf{x}),\bar{\mathcal{K}}_b(\textbf{y})\}&=-f_{ab}{}^c\phi_c{}^i\delta^2(\textbf{x}-\textbf{y}),\\
			\{\bar{\mathcal{K}}_a(\textbf{x}),\bar{\mathcal{K}}_b(\textbf{y})\}&=-f_{ab}{}^c\bar{\mathcal{K}}_c\delta^2(\textbf{x}-\textbf{y}),
		\end{align}
	\end{subequations}
which are the only non-vanishing Poisson brackets of the constraints $\phi_a{}^\mu$ and $\bar{\mathcal{K}}_a$. Hence $\phi_a{}^0$ and $\bar{\mathcal{K}}_a$ are first class constraints and $\phi_a{}^i$ are second class constraints. Thus we can use the second class constraints $\phi_a{}^i$ to restrict our phase space and define the corresponding Dirac bracket of the remaining canonical variables. In this case the only non-vanishing Dirac bracket of the dynamical fields is given by the following relation
	\begin{equation}
		\{A^a{}_i(\textbf{x}),A^b{}_j(\textbf{y})\}_{D.B}=\frac{2\pi}{k}g^{ab}\epsilon_{ij}\delta^2(\textbf{x}-\textbf{y}).
	\end{equation}
\subsection{Constructing the Gauge Generator}
As a next step we are interested in the generators that correspond to the gauge transformations induced by the first class constraints $\phi_a{}^0$ and $\bar{\mathcal{K}}_a$. A useful way to construct the generators is given by Castellani's algorithm \cite{Blagojevic:2002aa}. In the general case the gauge generator is given by
	\begin{equation}
		G=\lambda(t)G_0+\dot{\lambda}(t)G_1,
	\end{equation}
with $\dot{\lambda}(t)\equiv\frac{d\lambda(t)}{dt}$. The constraints $G_0$ and $G_1$ then have to fulfill the following relations
	\begin{subequations}
		\begin{align}
			G_1&=C_{PFC},\\
			G_0+\{G_1,\mathcal{H}_T\}&=C_{PFC},\\
			\{G_0,\mathcal{H}_T\}&=C_{PFC},
		\end{align}
	\end {subequations}
where $C_{PFC}$ denotes a primary first class constraint. These relations are fulfilled for $G_0=\bar{\mathcal{K}}_a$ and $G_1=\phi_a{}^0=\pi_a{}^0$. The smeared generator of gauge transformations has the following form
	\begin{equation}
		G[\lambda]=\int_\Sigma\extd^2x\left(D_0\lambda^a\pi_a{}^0+\lambda^a\bar{\mathcal{K}}_a\right).
	\end{equation}
Using \eqref{Intro:ConstraintBrackets} one can show by a straightforward calculation that this generator generates the following gauge transformations via $\delta_\lambda\bullet=\{\bullet,G[\lambda]\}$
	\begin{subequations}
		\begin{align}
			\delta_\lambda A^a{}_0&=D_0\lambda^a,\\
			\delta_\lambda A^a{}_i&=D_i\lambda^a,\\
			\delta_\lambda\pi_a{}^0&=-f_{ab}{}^c\lambda^b\pi_c{}^0,\\
			\delta_\lambda\pi_a{}^i&=\frac{k}{4\pi}\epsilon^{ij}g_{ab}\partial_j\lambda^b-f_{ab}{}^c\lambda^b\pi_c{}^i,\\
			\delta_\lambda\phi_a{}^i&=-f_{ab}{}^c\lambda^b\phi_c{}^i.
		\end{align}
	\end{subequations}
The generator $G$ that we have constructed via this method is only a preliminary result, since the presence of a boundary in our theory prevents  that the generator $G$ is properly functionally differentiable. We will fix this by first computing the full variation of the generator for a field independent gauge parameter $\lambda^a$
	\begin{align}\label{Intro:DeltaG}
		\delta G[\lambda]=&\int_\Sigma\extd^2x(\delta(D_0\lambda^a\pi_a{}^0)+\lambda^a\delta\bar{K}_a)=\nonumber\\
				      &\int_\Sigma\extd^2x\left(\dot{\lambda}^a\delta\pi_a{}^0-\lambda^af_{ab}{}^c(\delta A^b{}_0\pi_c{}^0+A^b{}_0\delta\pi_c{}^0)-\frac{k}{4\pi}
					\epsilon^{ij}g_{ab}\partial_j\lambda^a\delta A^b{}_i+\right.\nonumber\\
				      &\left.\partial_i\lambda^a\delta\pi_a{}^i-\lambda^af_{ab}{}^c(\delta A^b{}_i\pi_c{}^i+A^b{}_i\delta\pi_c{}^i)-
					\partial_i\left(\frac{k}{4\pi}\epsilon^{ij}g_{ab}\lambda^a\delta A^b{}_j+\lambda^a\delta\pi_a{}^i\right)\right)\nonumber\\
				   =&\int_\Sigma\extd^2x\left(f^a{}_{bc}\lambda^c\pi_a{}^\mu\delta A^b{}_\mu+D_\mu\lambda^a\delta\pi_a{}^\mu+
					\frac{k}{4\pi}\epsilon^{ij}g_{ab}\partial_i\lambda^a\delta A^b{}_j-\right.\nonumber\\
				      &\left.\partial_i\left(\frac{k}{4\pi}\epsilon^{ij}g_{ab}\lambda^a\delta A^b{}_j+\lambda^a\delta\pi_a{}^i\right)\right).
	\end{align}
The first three terms are regular bulk terms and thus do not spoil functional differentiability. The last term on the other hand is a boundary term that spoils functional differentiability. Thus in order to fix this one has to add a suitable boundary term to the gauge generator such that the variation of this additional boundary term cancels exactly the boundary term in  \eqref{Intro:DeltaG} i.e.
	\begin{equation}
		\delta\bar{G}[\lambda]=\delta G[\lambda]+\delta Q[\lambda],
	\end{equation}
with
	\begin{equation}
		\delta Q[\lambda]=\int_\Sigma\extd^2x\,\partial_i\left(\frac{k}{4\pi}\epsilon^{ij}g_{ab}\lambda^a\delta A^b{}_j+\lambda^a\delta\pi_a{}^i\right).
	\end{equation}
Setting the second class constraints $\phi_a{}^i\approx0$ strongly equal to zero, thus going into the reduced phase space and using in addition Stoke's theorem, the variation of the boundary charge can be written as
	\begin{equation}
		\delta Q[\lambda]=\frac{k}{2\pi}\int\extd\varphi g_{ab}\lambda^a\delta A^b{}_\varphi.
	\end{equation}
If we assume that the gauge parameter is field independent, then the boundary charge $Q[\lambda]$ is trivially integrable. This yields the following canonical boundary charge
	\begin{equation}
		Q[\lambda]=\frac{k}{2\pi}\int\extd\varphi g_{ab}\lambda^aA^b{}_\varphi.
	\end{equation}
\subsection{Partially Fixing the Gauge}
After performing the canonical analysis and having identified all the constraints we can turn our attention to an appropriate choice of gauge. Since we have found two first class constraints we are free to impose two sets of gauge conditions. One appropriate partial gauge fixing choice is given by \cite{Blagojevic:2002aa}
	\begin{subequations}\label{Intro:GaugeChoice}
		\begin{align}
			A_\rho={}&b^{-1}(\rho)\partial_\rho b(\rho),\\
			A_\varphi={}&b^{-1}(\rho)a_\varphi(\varphi,t)b(\rho),\\
			A_t={}&b^{-1}(\rho)a_t(\varphi,t)b(\rho),
		\end{align}
	\end{subequations}
with the group element
	\begin{equation}\label{Intro:BChoice}
		b(\rho)=e ^{\rho L_0}.
	\end{equation}
This choice of gauge automatically solves the flatness conditions $F_{t\rho}=0$ and $F_{\varphi\rho}=0$.

\clearpage

\section{AdS$_2\times\mathbb{R}$ and $\mathbb{H}_2\times\mathbb{R}$ for $\mathfrak{sl}(3)$}\label{AdSxR}
In this section we present appropriate boundary conditions on the connection $A$ with an AdS$_2\times\mathbb{R}$ or $\mathbb{H}_2\times\mathbb{R}$ background with $\mathbb{H}_2$ being the Lobachevsky plane. In order to construct such a background, an embedding that contains at least one singlet with $\textnormal{Tr}(S^2)\neq0$ is necessary. This leads to an embedding with three $\mathfrak{sl}(2)$ generators $L_n$ $(n=-1,0,1)$, two sets of generators $\psi_n^{\pm}$ $(n=-\frac{1}{2},\frac{1}{2})$ of spin $\frac{3}{2}$ and one singlet $S$ of spin 1. These generators fulfill the following commutation relations\footnote{For the matrix representations of these generators and the corresponding Killing form used for the computations in this section, please refer to Appendix (\ref{Appendix:NonPrincipalEmb}).}
	\begin{subequations}\label{AdSxR:CommRel}
		\begin{align}
			[L_n,L_m]={}&(n-m)L_{n+m},\label{AdSxR:CommRela}\\
			[L_n,S]={}&0,\label{AdSxR:CommRelb}\\
			[L_n,\psi_m^{\pm}]={}&(\frac{n}{2}-m)\psi_{n+m}^{\pm},\label{AdSxR:CommRelc}\\
			[S,\psi_m^{\pm}]={}&\pm\psi_m^{\pm},\label{AdSxR:CommReld}\\
			[\psi_n^{+},\psi_m^{-}]={}&L_{m+n}+\frac{3}{2}(m-n)S.\label{AdSxR:CommRele}
		\end{align}
	\end{subequations}
The existence of the two doublets in our representation allows us to consider linear combinations of the corresponding generators without spoiling \eqref{AdSxR:CommRelc}, and we used this freedom to define the $\psi_n^{\pm}$ in such a way that they are eigenstates of the adjoint action of the singlet $S$.\\
Consider the connections 
	\begin{subequations}\label{AdSxR:Connection}
		\begin{align}
			A_\rho={}&L_0&\bar{A}_\rho={}&-L_0\\
			A_\varphi={}&\sigma e^\rho L_1&\bar{A}_\varphi={}&-e^\rho L_{-1}\\
			A_t={}&0&\bar{A}_t={}&\sqrt{3}S
		\end{align}
	\end{subequations}
with some constant $\sigma=\pm1$. Using the following definition
	\begin{equation}
		g_{\mu\nu}:=\frac{1}{2}\textnormal{Tr}\left(A_\mu-\bar{A}_\mu\right)\left(A_\nu-\bar{A}_\nu\right),
	\end{equation} 
one obtains the following asymptotic background metric
	\begin{equation}
		\extd s^2=\extd t^2+\extd\rho^2-\sigma e^{2\rho}\extd\varphi^2. 
	\end{equation}
Depending on the sign of $\sigma$ this metric is asymptotically AdS$_2\times\mathbb{R}$ ($\sigma=1$) or $\mathbb{H}_2\times\mathbb{R}$ ($\sigma=-1$) with an Euclidean signature.
\subsection{Background Fluctuations and Boundary Conditions of the Connection}
Regarding the fluctuations of the background on the boundary which we assume to be located at $\rho\rightarrow\infty$, we consider as a starting point the following boundary conditions
	\begin{equation}\label{AdSxR:MetricBoundary}
		g_{\mu\nu}=\left(
			\begin{array}{ccc}
				1+\mathcal{O}(e^{-\rho})&\mathcal{O}(e^{-\rho})&\mathcal{O}(1)\\
				\cdot&1+\mathcal{O}(e^{-\rho})&\mathcal{O}(1)\\
				\cdot&\cdot&-\sigma e^{2\rho}+\mathcal{O}(e^\rho)
			\end{array}\right)_{\mu\nu},
	\end{equation}
where the coordinates are ordered as $t,\rho,\varphi$. These fluctuations are chosen in such a way that they agree with the asymptotic behavior of the first descendant of the  AdS$_2\times\mathbb{R}$ [$\mathbb{H}_2\times\mathbb{R}$] vacuum, which can be found in \cite{Grumiller:Ertl}. In fact it is even possible to have a bit stricter boundary conditions that still agree with the first descendant of the vacuum, which we will present in the following subsection. Since the structure of \eqref{AdSxR:Connection} suggests that $A$ and $\bar{A}$ are treated differently we will also state boundary conditions and the corresponding boundary charges for $A$ and $\bar{A}$ differently. A convenient notation for $A$ ($\bar{A}$) is given by the following splitting
	\begin{equation}\label{AdSxR:EasyNot}
		A_\mu=A^{(0)}_\mu+A^{(1)}_\mu,
	\end{equation} 
where $A^{(1)}_\mu$ will contain all the subleading parts that do not appear in the canonical boundary charge. Using the gauge choice given by \eqref{Intro:GaugeChoice} and \eqref{Intro:BChoice} one can write \eqref{AdSxR:EasyNot} as
	\begin{equation}
		A_\mu=b^{-1}a_\mu b=b^{-1}\left(a^{(0)}_\mu+a^{(1)}_\mu\right)b
	\end{equation}
and $\bar{A}_\mu$ as
	\begin{equation}
		\bar{A}_\mu=b\bar{a}_\mu b^{-1}=b\left(\bar{a}^{(0)}_\mu+\bar{a}^{(1)}_\mu\right)b^{-1}.
	\end{equation}
 the connection $A$ has to obey the following boundary conditions for $\rho\rightarrow\infty$ in order to fulfill \eqref{AdSxR:MetricBoundary}
	\begin{subequations}\label{AdSxR:BCs}
		\begin{align}
			a^{(0)}_\rho={}&L_0,\\
			a^{(0)}_\varphi={}&\sigma L_1+\frac{2\pi}{k}\left(-\mathcal{L}(\varphi)L_{-1}+\mathcal{W}^{+}_{\frac{1}{2}}(\varphi)\psi^{+}_{-\frac{1}{2}}
				-\mathcal{W}^{-}_{\frac{1}{2}}(\varphi)\psi^{-}_{-\frac{1}{2}}+\frac{3}{2}\mathcal{W}_0(\varphi)S\right),\\
			a^{(0)}_t={}&0,\\
			a^{(1)}_\mu={}&\mathcal{O}(e^{-2\rho}).
		\end{align}
	\end{subequations}
And for $\bar{A}$ the boundary conditions are given by
	\begin{subequations}\label{AdSxR:BCbs}
		\begin{align}
			\bar{a}^{(0)}_\rho={}&(-1+B(\varphi)e^{-\rho})L_0+\mathcal{O}(e^{-\rho})S,\label{AdSxR:BCbsa}\\
			\bar{a}^{(0)}_\varphi={}&\mathcal{O}(1)L_1+(-1+B(\varphi)e^{-\rho})L_{-1}+\mathcal{O}(1)\psi^{+}_{\frac{1}{2}}+
						\mathcal{O}(1)\psi^{-}_{\frac{1}{2}}-\frac{3\pi}{k}\bar{W}_0(\varphi)S,\label{AdSxR:BCbsb}\\
			\bar{a}^{(0)}_t={}&\left(\sqrt{3}+\mathcal{O}(e^{-\rho})\right)S,\\
			\bar{a}^{(1)}_\mu={}&\mathcal{O}(e^{-2\rho}).
		\end{align}
	\end{subequations}
We have chosen a normalization for the fields $\mathcal{L}$, $\mathcal{W}^{\pm}_{\frac{1}{2}}$, $\mathcal{W}_0$ and $\bar{\mathcal{W}}_0$ such that the corresponding canonical charge is conveniently normalized. The specific form and thus the appearance of the function $B(\varphi)$  as the subleading terms in \eqref{AdSxR:BCbsa} and \eqref{AdSxR:BCbsb} is a result of the requirement to fulfill the EOM asymptotically, i.e. $\left.F\right|_{\partial M}\rightarrow0$. If the subleading part of these two terms was not the same function, then the resulting connection would not be an asymptotically flat one. This is thus the only caveat if one tries to fulfill the boundary conditions \eqref{AdSxR:MetricBoundary}. The fluctuations appearing in $g_{\rho\rho}$ and $g_{\phi\phi}$ are not independent and are in fact identical. The requirement of asymptotical flatness of the connection also restricts the subleading terms of $A$ and $\bar{A}$ to be functions of only $\varphi$ and $\rho$.\\
As already mentioned, it is also possible to consider fluctuations of the metric that are a bit more restricted than \eqref{AdSxR:MetricBoundary}. Considering the fluctuations given by
	\begin{equation}\label{AdSxR:MetricBoundary2}
		g_{\mu\nu}=\left(
			\begin{array}{ccc}
				1+\mathcal{O}(e^{-2\rho})&\mathcal{O}(e^{-2\rho})&\mathcal{O}(1)\\
				\cdot&1+\mathcal{O}(e^{-2\rho})&\mathcal{O}(1)\\
				\cdot&\cdot&-\sigma e^{2\rho}+\mathcal{O}(1)
			\end{array}\right)_{\mu\nu},
	\end{equation}
then the corresponding connection $A$  has to obey the following boundary conditions
	\begin{subequations}\label{AdSxR:BCs1}
		\begin{align}
			a^{(0)}_\rho={}&L_0,\\
			a^{(0)}_\varphi={}&\sigma L_1+\frac{2\pi}{k}\left(-\mathcal{L}(\varphi)L_{-1}+\mathcal{W}^{+}_{\frac{1}{2}}(\varphi)\psi^{+}_{-\frac{1}{2}}
				-\mathcal{W}^{-}_{\frac{1}{2}}(\varphi)\psi^{-}_{-\frac{1}{2}}+\frac{3}{2}\mathcal{W}_0(\varphi)S\right),\label{AdSxR:BCs1b}\\
			a^{(0)}_t={}&0,\\
			a^{(1)}_\mu={}&\mathcal{O}(e^{-2\rho}).
		\end{align}
	\end{subequations}
For the $\bar{A}$-sector the connection has to obey asymptotically
	\begin{subequations}\label{AdSxR:BCbs1}
		\begin{align}
			\bar{a}^{(0)}_\rho={}&-L_0,\label{AdSxR:BCbsa1}\\
			\bar{a}^{(0)}_\varphi={}&\mathcal{O}(1)L_1-L_{-1}+\mathcal{O}(1)\psi^{+}_{\frac{1}{2}}+
						\mathcal{O}(1)\psi^{-}_{\frac{1}{2}}-\frac{3\pi}{k}\bar{W}_0(\varphi)S,\label{AdSxR:BCbsb1}\\
			\bar{a}^{(0)}_t={}&\sqrt{3}S,\\
			\bar{a}^{(1)}_\mu={}&\mathcal{O}(e^{-2\rho}).
		\end{align}
	\end{subequations}
The fluctuations resulting from \eqref{AdSxR:BCs1} and \eqref{AdSxR:BCbs1} obey \eqref{AdSxR:MetricBoundary2} and are completely arbitrary in contrast to the boundary conditions \eqref{AdSxR:BCbs} and \eqref{AdSxR:BCbs1}, which yielded fluctuations of the metric that had to be of a specific form. 
\subsection{Boundary Condition Preserving Gauge Transformations and Boundary Charges}
The boundary condition preserving gauge transformations and the resulting canonical boundary charges are the same for both boundary conditions presented, thus the following discussion applies to both cases. Since $A$ and $\bar{A}$ obey different boundary conditions, the corresponding boundary condition preserving gauge transformations and canonical boundary charges will be treated separately as well.\\
A gauge transformation with gauge parameter $\epsilon$ preserves a given set of boundary conditions if 
	\begin{equation}\label{AdSxR:BCPGT}
		\delta_\epsilon A^a{}_\mu=D_\mu\epsilon^a=\partial_\mu\epsilon^a+f^a{}_{bc}A^b{}_\mu\epsilon^c=\mathcal{O}\left(A^a{}_\mu\bigr|_{\partial\mathcal{M}}\right).
	\end{equation}
To be a little more specific on the notation: $A^a{}_\mu\bigr|_{\partial\mathcal{M}}$ denotes the subleading terms of the connection as $\rho\rightarrow\infty$. For the gauge choice \eqref{Intro:GaugeChoice} and the boundary conditions \eqref{AdSxR:BCs1b} this would mean for example that a boundary preserving gauge transformation has to satisfy
	\begin{gather}
		\delta_\epsilon A^{L_1}_\varphi=\mathcal{O}(e^{-\rho}),\quad\delta_\epsilon A^{L_0}_\varphi=\mathcal{O}(e^{-2\rho}),\quad 
		\delta_\epsilon A^{L_{-1}}_\varphi=\mathcal{O}(e^{-\rho}),\nonumber\\
		\delta_\epsilon A^{\psi^+_{\frac{1}{2}}}_\varphi=\mathcal{O}(e^{-\frac{3}{2}\rho}),\quad
		\delta_\epsilon A^{\psi^+_{-\frac{1}{2}}}_\varphi=\mathcal{O}(e^{-\frac{1}{2}\rho}),\nonumber\\
		\delta_\epsilon A^{\psi^-_{\frac{1}{2}}}_\varphi=\mathcal{O}(e^{-\frac{3}{2}\rho}),\quad
		\delta_\epsilon A^{\psi^-_{-\frac{1}{2}}}_\varphi=\mathcal{O}(e^{-\frac{1}{2}\rho}),\nonumber\\
		\delta_\epsilon A^{S}_\varphi=\mathcal{O}(1).
	\end{gather}
Using \eqref{AdSxR:BCPGT} one finds that the gauge transformations that preserve \eqref{AdSxR:BCs} [and \eqref{AdSxR:BCs1}] are given by
	\begin{equation}
		\hat{\epsilon}=b^{-1}\left(\epsilon^{(0)}+\epsilon^{(1)}\right)b.
	\end{equation}
The first part is given by
	\begin{equation}
		\epsilon^{(0)}=(\epsilon^1L_1+\epsilon^2L_0+\epsilon^3L_{-1}+\epsilon^4\psi^{+}_{\frac{1}{2}}+\epsilon^5\psi^{+}_{-\frac{1}{2}}+
			\epsilon^6\psi^{-}_{\frac{1}{2}}+\epsilon^7\psi^{-}_{-\frac{1}{2}}+\epsilon^8S)
	\end{equation}
with
	\begin{subequations}
		\begin{align}\label{AdSxR:BCPGTASectorDetail}
			\epsilon^1={}&\epsilon(\varphi),\quad\epsilon^2=-\frac{1}{\sigma}\epsilon'(\varphi),\\
			\epsilon^3={}&\frac{1}{2\sigma^2}\epsilon''(\varphi)-\frac{2\pi}{\sigma k}\left(\mathcal{L}(\varphi)\epsilon(\varphi)+\frac{1}{2}\left(
				\mathcal{W}^{-}_{\frac{1}{2}}(\varphi)\epsilon^{+}_{\frac{1}{2}}(\varphi)+
				\mathcal{W}^{+}_{\frac{1}{2}}(\varphi)\epsilon^{-}_{\frac{1}{2}}(\varphi)\right)\right),\\
			\epsilon^4={}&\epsilon^{+}_{\frac{1}{2}}(\varphi),\quad\epsilon^5=-\frac{1}{\sigma}\left({\epsilon^{+}_{\frac{1}{2}}}'(\varphi)-\frac{2\pi}{k}
				\left(\mathcal{W}^{+}_{\frac{1}{2}}(\varphi)\epsilon(\varphi)-\frac{3}{2}\mathcal{W}_0(\varphi)\epsilon^{+}_{\frac{1}{2}}(\varphi)\right)\right),\\
			\epsilon^6={}&\epsilon^{-}_{\frac{1}{2}}(\varphi),\quad\epsilon^7=-\frac{1}{\sigma}\left({\epsilon^{-}_{\frac{1}{2}}}'(\varphi)+\frac{2\pi}{k}
				\left(\mathcal{W}^{-}_{\frac{1}{2}}(\varphi)\epsilon(\varphi)-\frac{3}{2}\mathcal{W}_0(\varphi)\epsilon^{-}_{\frac{1}{2}}(\varphi)\right)\right),\\
			\epsilon^8={}&\epsilon_0(\varphi).
		\end{align}
	\end{subequations}
and the subleading parts are given by
	\begin{equation}
		\epsilon^{(1)}=\mathcal{O}(e^{-2\rho}).		
	\end{equation}
Having found the boundary condition preserving gauge transformations we are now interested how the fields $\mathcal{L}$, $\mathcal{W}^\pm_{\frac{1}{2}}$ and $\mathcal{W}_0$ transform under these gauge transformations. Since
	\begin{equation}
		A^{L_{-1}}_\varphi=-\frac{2\pi}{k}\mathcal{L}e^{-\rho},\quad A^{\psi^\pm_{-\frac{1}{2}}}_\varphi=\pm\frac{2\pi}{k}\mathcal{W}^\pm_{\frac{1}{2}}e^{-\frac{1}{2}\rho},
		\quad A^{S}_\varphi=\frac{3\pi}{k}\mathcal{W}_0,
	\end{equation}
it is easy to see that 
	\begin{equation}
		\delta A^{L_{-1}}_\varphi=-\frac{2\pi}{k}\delta\mathcal{L}e^{-\rho},\quad 
		\delta A^{\psi^\pm_{-\frac{1}{2}}}_\varphi=\pm\frac{2\pi}{k}\delta\mathcal{W}^\pm_{\frac{1}{2}}e^{-\frac{1}{2}\rho},
		\quad\delta A^{S}_\varphi=\frac{3\pi}{k}\delta\mathcal{W}_0,		
	\end{equation}
with $\delta=\delta_{\hat{\epsilon}}=\delta_{\epsilon_0}+\delta_{\epsilon}+\delta_{\epsilon^{+}_{\frac{1}{2}}}+\delta_{\epsilon^{-}_{\frac{1}{2}}}$. In order to find the correct boundary preserving gauge transformations we already calculated $\delta A^{L_{-1}}_\varphi$, $\delta A^{\psi^\pm_{-\frac{1}{2}}}_\varphi$ and $\delta A^{S}_\varphi$. Thus, one only has to look at the leading order contributions of these expressions and read off the transformation properties of the fields $\mathcal{L}$, $\mathcal{W}^\pm_{\frac{1}{2}}$ and $\mathcal{W}_0$. This leads to the following transformations 
	\begin{subequations}\label{AdSxR:Trafos}
		\begin{align}
			&\delta_{\epsilon_0}\mathcal{W}_0=\frac{k}{3\pi}\epsilon_0',\quad
			\delta_{\epsilon_0}\mathcal{W}^{\pm}_{\frac{1}{2}}=
			\mp\epsilon_0\mathcal{W}^{\pm}_{\frac{1}{2}},\quad
			\delta_{\epsilon_0}\mathcal{L}=0,\\
			&\delta_{\epsilon}\mathcal{W}_0=0,\quad
			\delta_{\epsilon}\mathcal{L}=-\frac{k}{4\pi}\epsilon'''+\sigma\left(2\epsilon'\mathcal{L}+
			\epsilon\mathcal{L}'\right),\nonumber\\
			&\delta_{\epsilon}\mathcal{W}^{\pm}_{\frac{1}{2}}=\sigma\left(\frac{3}{2}\epsilon'
			\mathcal{W}^{\pm}_{\frac{1}{2}}+\epsilon{\mathcal{W}^{\pm}_{\frac{1}{2}}}'
			\pm\frac{3\pi}{k}\epsilon\mathcal{W}^{\pm}_{\frac{1}{2}}\mathcal{W}_0\right),\label{AdSxR:TrafosEps}\\
			&\delta_{\epsilon^{\pm}_{\frac{1}{2}}}\mathcal{W}_0=\mp
			\mathcal{W}^{\mp}_{\frac{1}{2}}
			\epsilon_{\frac{1}{2}}^{\pm},\quad\delta_{\epsilon^{\pm}_{\frac{1}{2}}}\mathcal{W}^{\mp}_{\frac{1}{2}}=0,\nonumber\\
			&\delta_{\epsilon^{\pm}_{\frac{1}{2}}}\mathcal{W}^{\pm}_{\frac{1}{2}}=\pm\epsilon^{\pm}_{\frac{1}{2}}\mathcal{L}-
			\sigma\left(3{\epsilon^{\pm}_{\frac{1}{2}}}'\mathcal{W}_0+\frac{3}{2}\epsilon^{\pm}_{\frac{1}{2}}{\mathcal{W}_0}'\pm
			\frac{9\pi}{2k}\epsilon^{\pm}_{\frac{1}{2}}\mathcal{W}_0\mathcal{W}_0\pm\frac{k}{2\pi}{\epsilon^{\pm}_{\frac{1}{2}}}''\right)\nonumber\\
			&\delta_{\epsilon^{\pm}_{\frac{1}{2}}}\mathcal{L}=\frac{\sigma}{2}\left(
			\pm\frac{6\pi}{k}\mathcal{W}_0\mathcal{W}_{\frac{1}{2}}^{\mp}\epsilon_{\frac{1}{2}}^{\pm}+
			{\mathcal{W}_{\frac{1}{2}}^{\mp}}'\epsilon_{\frac{1}{2}}^{\mp}+3\mathcal{W}_{\frac{1}{2}}^{\mp}{\epsilon_{\frac{1}{2}}^{\pm}}'\right).
		\end{align}
	\end{subequations}
Please note that in order to obtain \eqref{AdSxR:Trafos} we used $\sigma^2=1$ in order to simplify some of the expressions. If $\sigma\neq\pm1$ then one has to replace $\sigma\rightarrow\frac{1}{\sigma}$ and $-\frac{k}{4\pi}\epsilon'''\rightarrow-\frac{k}{4\pi\sigma^2}\epsilon'''$ in \eqref{AdSxR:Trafos} and in all formulas appearing in section \ref{sec:ClassAsymptSymmAlgebra} to get the correct prefactors. The corresponding variation of the boundary charge is then given by
	\begin{equation}\label{AdSxR:BoundaryCharge}
		\delta\mathcal{Q}(\hat{\epsilon})=\int\extd\varphi\left(\delta\mathcal{L}\epsilon+\delta\mathcal{W}_0\epsilon_0+
		\delta\mathcal{W}_{\frac{1}{2}}^{+}\epsilon_{\frac{1}{2}}^{-}+\delta\mathcal{W}_{\frac{1}{2}}^{-}\epsilon_{\frac{1}{2}}^{+}\right).
	\end{equation}
Since the gauge parameters $\epsilon$, $\epsilon^\pm_{\frac{1}{2}}$, $\epsilon_0$ are field independent one can write the corresponding canonical charge as 
	\begin{equation}\label{AdSxR:BoundaryChargeFinal}
		\mathcal{Q}(\hat{\epsilon})=\int\extd\varphi\left(\mathcal{L}\epsilon+\mathcal{W}_0\epsilon_0+
		\mathcal{W}_{\frac{1}{2}}^{+}\epsilon_{\frac{1}{2}}^{-}+\mathcal{W}_{\frac{1}{2}}^{-}\epsilon_{\frac{1}{2}}^{+}\right).
	\end{equation}
Finding the boundary condition preserving gauge transformations and boundary charge for the $\bar{A}$-sector works exactly like for the $A$-sector. The only difference is that we have to preserve a different set of boundary conditions. The gauge transformations preserving \eqref{AdSxR:BCbs}  [and \eqref{AdSxR:BCbs1}]  are given by
	\begin{equation}\label{AdSxR:GaugeTrafosAbar}
		\bar{\epsilon}(\varphi)=b\left(\bar{\epsilon}^{(0)}+\bar{\epsilon}^{(1)}\right)b^{-1},
	\end{equation}
with
	\begin{equation}
		\bar{\epsilon}^{(0)}=\bar{\epsilon}_0(\varphi)S
	\end{equation}
and
	\begin{equation}
		\bar{\epsilon}^{(1)}=\mathcal{O}(e^{-2\rho}).
	\end{equation}
Comparing these gauge transformations with \eqref{AdSxR:BCPGTASectorDetail}, we see that the gauge transformations in the $\bar{A}$-sector are a lot more restricted. The reason for this is the presence of the singlet term at leading order in $\bar{A}^S_t$. Without this singlet term the boundary condition preserving gauge transformations for the $\bar{A}$-sector would look like \eqref{AdSxR:BCPGTASectorDetail}. This singlet term, however, is crucial in our discussion of AdS$_2\times\mathbb{R}$ [$\mathbb{H}_2\times\mathbb{R}$] since the singlet term generates the $\extd t^2$ part of the background. Thus, the same analysis as for the $A$-sector yields the following transformation
	\begin{equation}
		\delta_{\bar{\epsilon}_0}\bar{\mathcal{W}}_0=-\frac{k}{3\pi}{\bar{\epsilon}_0}'.
	\end{equation}
The corresponding variation of the boundary charge is given by
	\begin{equation}
		\delta\bar{\mathcal{Q}}(\bar{\epsilon})=\int\extd\varphi\delta\bar{\mathcal{W}}_0\bar{\epsilon}_0.
	\end{equation}
Again, the gauge parameters are field independent and thus the canonical charge for the $\bar{A}$-sector is given by
	\begin{equation}
		\bar{\mathcal{Q}}(\bar{\epsilon})=\int\extd\varphi\bar{\mathcal{W}}_0\bar{\epsilon}_0.
	\end{equation}
\subsection{Calculating the Classical Asymptotic Symmetry Algebra}\label{sec:ClassAsymptSymmAlgebra}
After computing \eqref{AdSxR:BCs} and \eqref{AdSxR:BCbs} one can calculate the Dirac brackets corresponding to the symmetry present at the boundary \cite{Campoleoni:2011hg}. The latter then yields the asymptotic symmetry algebra. Actually, there is a convenient short-cut that avoids the tedious calculation of Dirac brackets.
Given two fields $\mathcal{V}$, $\mathcal{W}$ and a canonical boundary charge $\hat{\mathcal{Q}}(\lambda)=\int\extd\varphi\,\lambda(\varphi)\mathcal{V}(\varphi)$ one can use
	\begin{equation}\label{AdSxR:DiracBracCalc}
		\delta_\lambda\mathcal{W}(\bar{\varphi})=-\{\hat{\mathcal{Q}}(\lambda),\mathcal{W}(\bar{\varphi})\}=-
		\int\extd\varphi\,\lambda(\varphi)\{\mathcal{V}(\varphi),\mathcal{W}(\bar{\varphi})\},
	\end{equation}
to determine $\{\mathcal{V}(\varphi),\mathcal{W}(\bar{\varphi})\}$, given that $\delta_\lambda\mathcal{W}(\bar{\varphi})$ has been calculated beforehand. The Dirac bracket $\{\mathcal{L}(\varphi),\mathcal{L}(\bar{\varphi})\}$ for example can be calculated via
	\begin{equation}\label{AdSxR:Vircalc}
		\delta_\epsilon\mathcal{L}(\bar{\varphi})=-\{\mathcal{Q}(\epsilon),\mathcal{L}(\Bar{\varphi})\}=-
		\int\extd\varphi\,\epsilon(\varphi)\{\mathcal{L}(\varphi),\mathcal{L}(\bar{\varphi})\}.
	\end{equation}
Equation \eqref{AdSxR:Vircalc} can be satisfied for
	\begin{equation}
		\{\mathcal{L}(\varphi),\mathcal{L}(\bar{\varphi})\}=-\frac{k}{4\pi}\delta'''(\varphi-\bar{\varphi})+
			\sigma\left(2\mathcal{L}(\bar{\varphi})\delta'(\varphi-\bar{\varphi})
			-\mathcal{L}'(\bar{\varphi})\delta(\varphi-\bar{\varphi})\right),
	\end{equation}
with $\delta'(\varphi-\bar{\varphi})=\partial_\varphi(\varphi-\bar{\varphi})$.
This can also be written in terms of $\delta_\epsilon\mathcal{L}$ as
	\begin{equation}\label{AdSxR:Condition}
		\{\mathcal{L}(\varphi),\mathcal{L}(\bar{\varphi})\}=-\delta_\epsilon\mathcal{L}(\bar{\varphi})
		\Bigr|_{\partial_{\bar{\varphi}}^n\epsilon(\bar{\varphi})=(-1)^n\partial_\varphi^n\delta(\varphi-\bar{\varphi})}.
	\end{equation}
Using \eqref{AdSxR:DiracBracCalc} this procedure can be repeated for all the other remaining fields and gauge parameters appearing in \eqref{AdSxR:Trafos}. This yields the following Dirac brackets with the convention that all fields appearing on the right hand side depend on $\bar{\varphi}$ and $\delta'(\varphi-\bar{\varphi})\equiv\partial_{\varphi}\delta(\varphi-\bar{\varphi})$.
	\begin{subequations}\label{AdSxR:W32AlgebraDirac}
		\begin{align}
			&\{\mathcal{W}_0(\varphi),\mathcal{W}_0(\bar{\varphi})\}=\frac{k}{3\pi}\delta'(\varphi-\bar{\varphi}),\\
			&\{\mathcal{W}_0(\varphi),\mathcal{L}(\bar{\varphi})\}=0,\label{AdSxR:W32AlgebraDiracb}\\
			&\{\mathcal{W}_0(\varphi),\mathcal{W}_{\frac{1}{2}}^{\pm}(\bar{\varphi})\}=
			\pm\mathcal{W}_{\frac{1}{2}}^{\pm}\delta(\varphi-\bar{\varphi}),\\
			&\{\mathcal{L}(\varphi),\mathcal{L}(\bar{\varphi})\}=-\frac{k}{4\pi}\delta'''(\varphi-\bar{\varphi})+
			\sigma\left(2\mathcal{L}\delta'(\varphi-\bar{\varphi})
			-\mathcal{L}'\delta(\varphi-\bar{\varphi})\right),\label{AdSxR:W32AlgebraDiracLL}\\
			&\{\mathcal{L}(\varphi),\mathcal{W}_{\frac{1}{2}}^{\pm}(\bar{\varphi})\}=
			\sigma\left(\frac{3}{2}\mathcal{W}_{\frac{1}{2}}^{\pm}\delta'(\varphi-\bar{\varphi})-
			{\mathcal{W}_{\frac{1}{2}}^{\pm}}'\delta(\varphi-\bar{\varphi})\mp\frac{3\pi}{k}\mathcal{W}_{\frac{1}{2}}^{\pm}
			\mathcal{W}_0\delta(\varphi-\bar{\varphi})\right),\label{AdSxR:W32AlgebraDiracd}\\
			&\{\mathcal{W}_{\frac{1}{2}}^{+}(\varphi),\mathcal{W}_{\frac{1}{2}}^{-}(\bar{\varphi})\}=
			\mathcal{L}\delta(\varphi-\bar{\varphi})+\sigma\left(-3\mathcal{W}_0\delta'(\varphi-\bar{\varphi})+
			\frac{3}{2}{\mathcal{W}_0}'\delta(\varphi-\bar{\varphi})-\right.\nonumber\\
			&\qquad\qquad\qquad\qquad\quad
			\left.
			\frac{9\pi}{2k}\mathcal{W}_0\mathcal{W}_0\delta(\varphi-\bar{\varphi})-
			\frac{k}{2\pi}\delta''(\varphi-\bar{\varphi})\right)\\
			&\{\mathcal{W}_{\frac{1}{2}}^{+}(\varphi),\mathcal{W}_{\frac{1}{2}}^{+}(\bar{\varphi})\}=
			\{\mathcal{W}_{\frac{1}{2}}^{-}(\varphi),\mathcal{W}_{\frac{1}{2}}^{-}(\bar{\varphi})\}=0.
		\end{align}
	\end{subequations}
This algebra is written in a non-primary basis, as one can see by looking at \eqref{AdSxR:W32AlgebraDiracb} and \eqref{AdSxR:W32AlgebraDiracd}. This can be fixed by a shift of $\mathcal{L}$ given by
	\begin{equation}
		\mathcal{L}\rightarrow\mathcal{L}+\frac{3\pi\sigma}{2k}\mathcal{W}_0\mathcal{W}_0\equiv\hat{\mathcal{L}}.
	\end{equation}
After applying this shift, the non-vanishing Dirac brackets for the $A$-sector are given by
	\begin{subequations}
		\begin{align}
			&\{\mathcal{W}_0(\varphi),\mathcal{W}_0(\bar{\varphi})\}=\frac{k}{3\pi}\delta'(\varphi-\bar{\varphi}),\\
			&\{\mathcal{W}_0(\varphi),\hat{\mathcal{L}}(\bar{\varphi})\}=\sigma\mathcal{W}_0\delta'(\varphi-\bar{\varphi}),\\
			&\{\mathcal{W}_0(\varphi),\mathcal{W}_{\frac{1}{2}}^{\pm}(\bar{\varphi})\}=
			\pm\mathcal{W}_{\frac{1}{2}}^{\pm}\delta(\varphi-\bar{\varphi}),\\
			&\{\hat{\mathcal{L}}(\varphi),\hat{\mathcal{L}}(\bar{\varphi})\}=-\frac{k}{4\pi}\delta'''(\varphi-\bar{\varphi})+
			\sigma\left(2\hat{\mathcal{L}}\delta'(\varphi-\bar{\varphi})
			-\hat{\mathcal{L}}'\delta(\varphi-\bar{\varphi})\right),\\
			&\{\hat{\mathcal{L}}(\varphi),\mathcal{W}_{\frac{1}{2}}^{\pm}(\bar{\varphi})\}=
			\sigma\left(\frac{3}{2}\mathcal{W}_{\frac{1}{2}}^{\pm}\delta'(\varphi-\bar{\varphi})-
			{\mathcal{W}_{\frac{1}{2}}^{\pm}}'\delta(\varphi-\bar{\varphi})\right),\\
			&\{\mathcal{W}_{\frac{1}{2}}^{+}(\varphi),\mathcal{W}_{\frac{1}{2}}^{-}(\bar{\varphi})\}=
			\hat{\mathcal{L}}\delta(\varphi-\bar{\varphi})+\sigma\left(-3\mathcal{W}_0\delta'(\varphi-\bar{\varphi})+
			\frac{3}{2}{\mathcal{W}_0}'\delta(\varphi-\bar{\varphi})-\right.\nonumber\\
			&\qquad\qquad\qquad\qquad\quad
			\left.
			\frac{6\pi}{k}\mathcal{W}_0\mathcal{W}_0\delta(\varphi-\bar{\varphi})-
			\frac{k}{2\pi}\delta''(\varphi-\bar{\varphi})\right).
		\end{align}
	\end{subequations}
The Dirac brackets for the $\bar{A}$-sector are given by
	\begin{equation}
		\{\bar{\mathcal{W}}_0(\varphi),\bar{\mathcal{W}}_0(\bar{\varphi})\}=-\frac{k}{3\pi}\delta'(\varphi-\bar{\varphi}).
	\end{equation}
Thus, the boundary conditions \eqref{AdSxR:BCs} and \eqref{AdSxR:BCbs} give rise to one copy of a classical $\mathcal{W}_3^{(2)}$ algebra, which is also called Polyakov-Bershadsky algebra for the $A$-sector and a $\mathfrak{u}(1)$ current algebra with a central extension for the $\bar{A}$-sector. Therefore, we obtain $\mathcal{W}_3^{(2)}\oplus\mathfrak{u}(1)$ as the asymptotic symmetry algebra.
Since we are interested in the central charges of the corresponding boundary theory, we will first express \eqref{AdSxR:W32AlgebraDirac} in terms of its Fourier modes but without taking normal ordering into account and thus obtain the "classical central charges" of the boundary theory. After obtaining this algebra and replacing the Dirac brackets with commutators, we will focus on normal ordering issues and determine the effective central charge of the boundary theory. Since the central terms in \eqref{AdSxR:W32AlgebraDirac} can be rescaled arbitrarily and thus the central charges are not unique, one has to find a way to fix this.
In the case we are considering, at least in the $A$-sector one has the additional information of the algebra without a central extension given by \eqref{AdSxR:CommRel}. Thus if one rescales the relations \eqref{AdSxR:W32AlgebraDirac} in such a way that the corresponding non-centrally extended part of the commutator algebra agrees with \eqref{AdSxR:CommRel}, then the correct central charges can be read off directly. Using the following mode expansion 
	\begin{equation}\label{AdSxR:Delta}
		\mathcal{L}(\varphi)=\frac{\sigma}{2\pi}\sum_{n\in\mathbb{Z}}L_ne^{-in\varphi}\quad\textnormal{and}
		\quad\delta(\varphi-\bar{\varphi})=\frac{1}{2\pi}\sum_{n\in\mathbb{Z}}e^{-in(\varphi-\bar{\varphi})},
	\end{equation}
where we have shifted the zero mode as
	\begin{equation}
		L_0\rightarrow L_0-\frac{k}{4}\delta_{n,0},
	\end{equation}
and plugging this expansion into \eqref{AdSxR:W32AlgebraDiracLL} one obtains
	\begin{align}
		\sum_{n,m\in\mathbb{Z}}e^{-i(n\varphi+m\bar{\varphi})}\{L_n,L_m\}={}&\sum_{n,p\in\mathbb{Z}}
		e^{-ip\bar{\varphi}-in(\varphi-\bar{\varphi})}\left(L_p-\frac{k}{4}\delta_{p,0}\right)(-2in+ip)\nonumber\\
		{}&-\frac{k}{2}\sum_{n\in\mathbb{Z}}in^3e^{-in(\varphi-\bar{\varphi})}.
	\end{align}
Making an appropriate shift $p-n=m$ and using the orthogonality property of the complex exponential function $\int\extd\varphi e^{i(n-m)\varphi}=2\pi\delta_{n,m}$ one obtains
	\begin{equation}
		\{L_n,L_m\}=i(m-n)L_{m+n}-i\frac{k}{2}n(n^2-1)\delta_{m+n,0}.
	\end{equation}
Replacing the Dirac bracket with a commutator using $i\{\cdot,\cdot\}\rightarrow[\cdot,\cdot]$ one obtains 
	\begin{equation}
		[L_n,L_m]=(n-m)L_{m+n}+\frac{c}{12}n(n^2-1)\delta_{m+n,0}.
	\end{equation}
with the central charge $c=6k$. Using the same procedure with all remaining Dirac brackets \eqref{AdSxR:W32AlgebraDirac} and the following mode expansions
	\begin{equation}
		\mathcal{W}_0(\varphi)=\frac{i}{2\pi}\sum_{n\in\mathbb{Z}}J_ne^{-in\varphi}\quad\textnormal{and}\quad
		\mathcal{W}_{\frac{1}{2}}^{\pm}(\varphi)=\frac{(i\sigma)^{\frac{1\mp1}{2}}}{2\pi}\sum_{n\in\mathbb{Z}+\frac{1}{2}}G_n^{\pm}e^{-in\varphi},
	\end{equation}
one obtains the following (classical) commutation relations
	\begin{subequations}\label{AdSxR:W32Algebra}
		\begin{align}
			&[J_n,J_m]=-\frac{2k}{3}n\delta_{n+m,0},\\
			&[J_n,L_m]=0,\\
			&[J_n,G_m^{\pm}]=\pm G_{m+n}^{\pm},\\
			&[L_n,L_m]=(n-m)L_{m+n}+\frac{c}{12}n(n^2-1)\delta_{n+m,0},\\
			&[L_n,G_m^{\pm}]=\left(\frac{n}{2}-m\right)G_{n+m}^{\pm}\pm\frac{3}{4k}\sum_{p\in\mathbb{Z}}(G_{m+n-p}^{\pm}J_p+J_pG_{m+n-p}^{\pm}),\\
			&[G_n^{+},G_m^{-}]=L_{m+n}+\frac{3}{2}(m-n)J_{m+n}+\frac{9}{4k}\sum_{p\in\mathbb{Z}}J_{m+n-p}J_p+k(n^2-\frac{1}{4})\delta_{m+n,0},\label{AdSxR:W32Algebra6}\\
			&[G_n^{+},G_m^{+}]=[G_n^{-},G_m^{-}]=0.
		\end{align}
	\end{subequations}
Please note that in order to calculate \eqref{AdSxR:W32Algebra6} the following definition for $\delta(\varphi-\bar{\varphi})$ has been used
	\begin{equation}\label{AdSxR:Delta1}
		\delta(\varphi-\bar{\varphi})=\frac{1}{2\pi}\sum_{n\in\mathbb{Z}+\frac{1}{2}}e^{-in(\varphi-\bar{\varphi})}.
	\end{equation}
This definition is necessary in order to satisfy
	\begin{equation}
		\int\extd\varphi\epsilon^{\pm}_{\frac{1}{2}}(\varphi)\delta(\varphi-\bar{\varphi})=\epsilon^{\pm}_{\frac{1}{2}}(\bar{\varphi}),
	\end{equation}
with
	\begin{equation}
		\epsilon^{\pm}_{\frac{1}{2}}(\varphi)=\frac{1}{2\pi}\sum_{n\in\mathbb{Z}+\frac{1}{2}}\epsilon^{\pm}_ne^{-in\varphi}.
	\end{equation}
Thus if one tries to write a Dirac bracket in terms of Fourier modes that has been obtained by satisfying 
	\begin{equation}
		\delta_{\epsilon^{\pm}_{\frac{1}{2}}}\mathcal{W}^{\pm}_{\frac{1}{2}}(\bar{\varphi})=-
		\int\extd\varphi\epsilon^{\pm}_{\frac{1}{2}}(\varphi)\{\mathcal{W}^{\mp}_{\frac{1}{2}}(\varphi),\mathcal{W}^{\pm}_{\frac{1}{2}}(\bar{\varphi})\},
	\end{equation}
one has to use \eqref{AdSxR:Delta1} rather than \eqref{AdSxR:Delta}. In addition, we used
	\begin{equation}
		(\mathcal{W}_{\frac{1}{2}}^{\pm}\mathcal{W}_0)(\bar{\varphi})=\frac{1}{2}(\mathcal{W}_{\frac{1}{2}}^{\pm}\mathcal{W}_0+
		\mathcal{W}_0\mathcal{W}_{\frac{1}{2}}^{\pm})(\bar{\varphi}).
	\end{equation}
The algebraic relations \eqref{AdSxR:W32Algebra} can again be brought into a form where all fields appearing are proper Virasoro primaries. This is a similar shift to the one done in the case of the Dirac bracket algebra and is given by
	\begin{equation}
		L_n\rightarrow\hat{L}_n\equiv L_n-\frac{3}{4k}\sum_{p\in\mathbb{Z}}J_{n-p}J_p.
	\end{equation} 	
This yields the following algebra
	\begin{subequations}\label{AdSxR:W32AlgebraShifted}
		\begin{align}
			&[J_n,J_m]=-\frac{2k}{3}n\delta_{n+m,0},\\
			&[J_n,\hat{L}_m]=nJ_{n+m},\\
			&[J_n,G_m^{\pm}]=\pm G_{m+n}^{\pm},\\
			&[\hat{L}_n,\hat{L}_m]=(n-m)\hat{L}_{m+n}+\frac{c}{12}n(n^2-1)\delta_{n+m,0},\\
			&[\hat{L}_n,G_m^{\pm}]=\left(\frac{n}{2}-m\right)G_{n+m}^{\pm},\\
			&[G_n^{+},G_m^{-}]=\hat{L}_{m+n}+\frac{3}{2}(m-n)J_{m+n}+\frac{3}{k}\sum_{p\in\mathbb{Z}}J_{m+n-p}J_p+k(n^2-\frac{1}{4})\delta_{m+n,0},\label{AdSxR:W32AlgebraShifted6}\\
			&[G_n^{+},G_m^{+}]=[G_n^{-},G_m^{-}]=0.
		\end{align}
	\end{subequations}
\subsection{Quantum $\mathcal{W}_3^{(2)}$ and $\mathfrak{u}(1)$ Current Algebra}
Since we are interested in the quantum mechanical version of \eqref{AdSxR:W32Algebra}, we also have to take normal ordering into account whenever products of Fourier modes appear. 
The symbol $:\,:$ denotes normal ordering which we defined as follows
	\begin{equation}
		\sum_{p\in\mathbb{Z}}:J_{n-p}J_p:=\sum_{p\geq0}J_{n-p}J_p+\sum_{p<0}J_pJ_{n-p}.
	\end{equation}
However, since the algebraic relations \eqref{AdSxR:W32AlgebraShifted} are given in terms of large $c$ or equivalently large $k$, it is possible that all coefficients that contain factors of $k$ obtain quantum corrections of $\mathcal{O}(1)$. Thus when introducing normal ordering these corrections have to be determined. Possibly the easiest way to do this is to consider the following algebra
	\begin{subequations}\label{AdSxR:W32AlgebraQuantum}
		\begin{align}
			&[J_n,J_m]=C_1n\delta_{n+m,0},\\
			&[J_n,\hat{L}_m]=nJ_{n+m},\\
			&[J_n,G_m^{\pm}]=\pm G_{m+n}^{\pm},\\
			&[\hat{L}_n,\hat{L}_m]=(n-m)\hat{L}_{m+n}+\frac{\hat{c}}{12}n(n^2-1)\delta_{n+m,0},\\
			&[\hat{L}_n,G_m^{\pm}]=\left(\frac{n}{2}-m\right)G_{n+m}^{\pm},\\
			&[G_n^{+},G_m^{-}]=C_2\hat{L}_{m+n}+\frac{3}{2}C_3(m-n)J_{m+n}+C_4\sum_{p\in\mathbb{Z}}:J_{m+n-p}J_p:+C_5(n^2-\frac{1}{4})\delta_{m+n,0},\label{AdSxR:W32AlgebraQuantum6}\\
			&[G_n^{+},G_m^{+}]=[G_n^{-},G_m^{-}]=0,
		\end{align}
	\end{subequations}
and calculate the Jacobi identities which yield relations between these coefficients that allow us to fix them such that we have a consistent algebra (at least consistent with respect to the Jacobi identities). Even though the coefficients $C_2$ and $C_3$ are equal to 1 in \eqref{AdSxR:W32AlgebraShifted} and do not contain factors of $k$, a rescaling of $G_n^{\pm}$ by a factor of $\sqrt{k}$ could easily produce such a $k$ dependence and thus one has to consider these two coefficients not to be fixed to $1$. In addition, we fixed the shift of the normal ordered Virasoro modes to be
	\begin{equation}\label{AdSxR:NormOrdShiftVir}
		L_n\rightarrow\hat{L}_n\equiv L_n+\frac{1}{2C_1}\sum_{p\in\mathbb{Z}}:J_{n-p}J_p:.
	\end{equation}
This normalization ensures that normal ordering of the Virasoro modes results in the expected shift of the central charge $c=6k$ by $+1$, thus yielding a preliminary shift of the central charge $c\rightarrow c+1$. This shift, however, will also be further modified once the Jacobi identities have to be satisfied. Calculating the Jacobi identities yields the following relations between the coefficients
	\begin{subequations}\label{AdSxR:JacobiRel}
		\begin{align}
			&C_2+3C_3+2C_1C_4=0, &C_5+\frac{3}{2}C_3C_1=0,\\
			&C_2\hat{c}-6C_5+2C_1C_4=0, &C_3-C_2-\frac{2}{3}C_4=0.
		\end{align}
	\end{subequations}
Since we have four equations but six free parameters, we have the freedom to fix two of them and the remaining four coefficients are then determined by the relations \eqref{AdSxR:JacobiRel}. Since we already know what the algebra looks like in the classical case for large $k$, one viable choice of coefficients would be
	\begin{equation}
		C_1=-\frac{2k}{3},\quad C_2=1.
	\end{equation}
This yields the following coefficients and shifted central charge
	\begin{subequations}
		\begin{align}
			C_3=1+\frac{8}{4k-6},\quad &C_4=\frac{12}{4k-6},\quad C_5=k\left(1+\frac{8}{4k-6}\right),\\
			&\hat{c}=\frac{32k}{2k-3}+6k.
		\end{align}
	\end{subequations}
This is essentially the quantum $\mathcal{W}_3^{(2)}$ algebra found by Polyakov and Bershadsky in \cite{Bershadsky:1990bg,Polyakov:1989dm} but with a different $k$ and different normalization of the spin-$\frac{3}{2}$ modes. In order to bring this algebra in a more familiar form, we apply the following shift of $k$ and renormalization of $G_n^{\pm}$
	\begin{equation}\label{AdSxR:KGShift}
		k\rightarrow-\left(\hat{k}+\frac{3}{2}\right),\quad G_n^{\pm}\sqrt{-(\hat{k}+3)}\rightarrow\hat{G}_n^{\pm}.
	\end{equation}
This results in the following algebra
	\begin{subequations}\label{AdSxR:W32AlgebraQuantumBershadsky}
		\begin{align}
			&[J_n,J_m]=\frac{2\hat{k}+3}{3}n\delta_{n+m,0},\label{AdSxR:W32AlgebraQuantumBershadsky1}\\
			&[J_n,\hat{L}_m]=nJ_{n+m},\\
			&[J_n,\hat{G}_m^{\pm}]=\pm\hat{G}_{m+n}^{\pm},\\
			&[\hat{L}_n,\hat{L}_m]=(n-m)\hat{L}_{m+n}+\frac{\hat{c}}{12}n(n^2-1)\delta_{n+m,0},\\
			&[\hat{L}_n,\hat{G}_m^{\pm}]=\left(\frac{n}{2}-m\right)\hat{G}_{n+m}^{\pm},\\
			&[\hat{G}_n^{+},\hat{G}_m^{-}]=-(\hat{k}+3)\hat{L}_{m+n}+\frac{3}{2}(\hat{k}+1)(n-m)J_{m+n}
			+3\sum_{p\in\mathbb{Z}}:J_{m+n-p}J_p:+\nonumber\\
			&\qquad\qquad\quad\quad \frac{(\hat{k}+1)(2\hat{k}+3)}{2}(n^2-\frac{1}{4})\delta_{m+n,0},\label{AdSxR:W32AlgebraQuantumBershadsky6}\\
			&[\hat{G}_n^{+},\hat{G}_m^{+}]=[\hat{G}_n^{-},\hat{G}_m^{-}]=0,
		\end{align}
	\end{subequations}
with
	\begin{equation}
		\hat{c}=25-\frac{24}{\hat{k}+3}-6(\hat{k}+3)=-\frac{(2\hat{k}+3)(3\hat{k}+1)}{\hat{k}+3}.
	\end{equation}
It is easy to see that the central charge $\hat{c}$ is only non-negative for a small range of $\hat{k}$, which is given by the interval $-\frac{1}{3}\leq\hat{k}\leq-\frac{3}{2}$. The maximum value of the central charge is $\hat{c}=1$, which is obtained for $\hat{k}=-1$. Thus, it is not possible to obtain a unitary field theory dual of AdS$_2\times\mathbb{R}$ or $\mathbb{H}_2\times\mathbb{R}$ in the semi-classical limit $|\hat{k}|\rightarrow\infty$ \cite{Castro:2012bc}.\\
In the $\bar{A}$-sector we have only one Poisson bracket corresponding to a $\mathfrak{u}(1)$ current algebra with a central extension. The value of the central charge corresponding to this central extension is not unique, since we can always rescale the fields or corresponding Fourier modes. Thus, with the following mode expansion
	\begin{equation}
		\bar{\mathcal{W}_0}(\varphi)=\frac{1}{2\pi}\sum_{n\in\mathbb{Z}}\bar{J}_ne^{-in\varphi},
	\end{equation}
and the same shift of the Chern-Simons level $k$ as in \eqref{AdSxR:KGShift}, one obtains the following commutator algebra
	\begin{equation}
		[\bar{J}_n,\bar{J}_m]=\frac{2\hat{k}+3}{3}n\delta_{n+m,0}.
	\end{equation}
\subsection{Unitarity of the Resulting CFT}
Having found the asymptotic symmetry algebra we are now interested if it is possible to obtain a unitary CFT for certain values of $\hat{k}$ or not. Thus, we have to check if there are any unphysical states i.e. states with negative norm present. Since the $\bar{A}$-sector only consists of a $\mathfrak{u}(1)$ algebra and it is not hard to find unitary representations for this algebra, we will focus with our analysis on the $A$-sector containing the $\mathcal{W}_3^{(2)}$ algebra where the existence of unitary representations is not obvious at first glance.\\
Let $|a;N\rangle$, with $a=1,\ldots,N$ denote a basis of states for a given level $N$. Then any state at level $N$ can be written as the following linear combination
	\begin{equation}
		|\psi;N\rangle=\sum_{a=1}^N\lambda_a|a;N\rangle,
	\end{equation}
where $\lambda_a\in\mathbb{C}$ are some arbitrary constants. The norm of such a state is then given by
	\begin{equation}
		\langle\psi;N|\psi;N\rangle=\sum_{a,b}^N\lambda^\dagger_a\underbrace{\langle a;N|b;N\rangle}_{K^{(N)}_{ab}}\lambda_b,
	\end{equation}
where $K^{(N)}_{ab}$ denotes the Gramian matrix at level $N$. Thus, in order to have a unitary theory the Gramian matrix has to be positive semidefinite.\\
Since all modes appearing in the $\mathcal{W}_3^{(2)}$ algebra are proper Virasoro primaries, their action on the vacuum state is given by
	\begin{equation}\label{AdSxR:AnnihilationWeight}
		L_n|0\rangle=0,\quad J_n|0\rangle=0,\quad \hat{G}^\pm_n|0\rangle=0\quad\textnormal{for}\quad n>-h_i
	\end{equation} 
where $h_i$ (with $i=L,J,\hat{G}$) denotes the conformal weight of the primary fields that correspond to the given modes $L_n$, $J_n$ and $\hat{G}^\pm_n$ respectively. The hermitian conjugate of the modes is defined as
	\begin{equation}\label{AdSxR:AllConjugate}
		\left(L_n\right)^\dagger\equiv L_{-n},\quad\left(J_n\right)^\dagger\equiv J_{-n},\quad\left(\hat{G}^\pm_n\right)^\dagger\equiv \hat{G}^\mp_{-n}.
	\end{equation}
The hermitian conjugate of $\hat{G}^\pm_n$ may look strange, but is in fact a direct consequence of the quantum $\mathcal{W}^{(2)}_3$ algebra. After defining $\left(L_n\right)^\dagger$ and $\left(J_n\right)^\dagger$ one can look at
	\begin{align}
		\left(\left[\hat{G}_n^{+},\hat{G}_m^{-}\right]\right)^\dagger=&\left[\left(\hat{G}^-_m\right)^\dagger,
		\left(\hat{G}^+_n\right)^\dagger\right]=\nonumber\\
		=&-(\hat{k}+3)\hat{L}^\dagger_{m+n}+\frac{3}{2}(\hat{k}+1)(n-m)J^\dagger_{m+n}
			+3\sum_{p\in\mathbb{Z}}\left(:J_{m+n-p}J_p:\right)^\dagger+\nonumber\\
			&\qquad\qquad\quad\quad \frac{(\hat{k}+1)(2\hat{k}+3)}{2}(n^2-\frac{1}{4})\delta_{m+n,0}.
	\end{align}
After using $\left(L_n\right)^\dagger\equiv L_{-n}$ and $\left(J_n\right)^\dagger\equiv J_{-n}$ one obtains the following
		\begin{align}
		\left[\left(\hat{G}^-_m\right)^\dagger,\left(\hat{G}^+_n\right)^\dagger\right]
		=&-(\hat{k}+3)\hat{L}_{-(m+n)}+\frac{3}{2}(\hat{k}+1)(n-m)J_{-(m+n)}
			+3\sum_{p\in\mathbb{Z}}:J_{-(m+n)-p}J_p:+\nonumber\\
			&\frac{(\hat{k}+1)(2\hat{k}+3)}{2}(n^2-\frac{1}{4})\delta_{m+n,0}=
		\left[\hat{G}_{-m}^{+},\hat{G}_{-n}^{-}\right].
	\end{align}
Thus, in general the hermitean conjugate of $\hat{G}^\pm_n$ is given by
	\begin{equation}
		\left(\hat{G}^\pm_n\right)^\dagger\equiv(\alpha)^{\pm1}\hat{G}^\mp_{-n},\label{AdSxR:GConjugate}
	\end{equation}
were $\alpha$ could in principle be any complex number. Since in any quantum field theory the n-point correlation functions have to be real valued functions, we get an additional restriction on $\alpha$. Consider for example the norm of the state $\hat{G}^+_{-\frac{3}{2}}|0\rangle$. Reality of the norm requires
	\begin{equation}
		\langle0|\left(\hat{G}^+_{-\frac{3}{2}}\right)^\dagger\hat{G}^+_{-\frac{3}{2}}|0\rangle=
		\left(\langle0|\left(\hat{G}^+_{-\frac{3}{2}}\right)^\dagger
		\hat{G}^+_{-\frac{3}{2}}|0\rangle\right)^\dagger.
	\end{equation}
Using \eqref{AdSxR:GConjugate} one gets
	\begin{equation}
		\alpha\langle0|\hat{G}^-_{\frac{3}{2}}\hat{G}^+_{-\frac{3}{2}}|0\rangle=\alpha^\ast
		\langle0|\hat{G}^-_{\frac{3}{2}}\hat{G}^+_{-\frac{3}{2}}|0\rangle
	\end{equation}
and thus
	\begin{equation}
		\alpha=\alpha^\ast.
	\end{equation}
Hence without loss of generality we can set $\alpha=1$ and arrive at the relations given by \eqref{AdSxR:AllConjugate}. Having properly defined the hermitean conjugate of the modes present in the $\mathcal{W}^{(2)}_3$ algebra, one can look for possible negative norm states on the first levels of the resulting CFT.\\
On level 1 there is only the state $J_{-1}|0\rangle$ present. In this case the Gramian matrix is simply the norm of the state and is given by
	\begin{equation}\label{AdSxR:Level1Norm}
		\langle0|J_1J_{-1}|0\rangle=\frac{2\hat{k}+3}{3},
	\end{equation}
which is non-negative for $\hat{k}\geq-\frac{3}{2}$.
The level $\frac{3}{2 }$ contains two states $\hat{G}^+_{-\frac{3}{2}}|0\rangle$ and $\hat{G}^-_{-\frac{3}{2}}|0\rangle$. Thus, the Gramian matrix at level $\frac{3}{2}$ is given by
	\begin{equation}\label{AdSxR:LevelOnehalfNorm}
		K^{(\frac{3}{2})}=(\hat{k}+1)(2\hat{k}+3)\left(
		\begin{array}{cc}
			-1&0\\
			0&1					
		\end{array}\right),
	\end{equation}
with the basis vectors arranged as $\hat{G}^+_{-\frac{3}{2}}|0\rangle$, $\hat{G}^-_{-\frac{3}{2}}|0\rangle$. At this level we encounter a crucial difference to the similar and maybe more familiar $\mathcal{N}=2$ superconformal algebra \cite{Ademollo:1975an}. In the case of the $\mathcal{W}^{(2)}_3$ algebra where the modes $\hat{G}^+_n$ and $\hat{G}^-_n$ obey commutation rather than anticommutation relations the norm of the states $\hat{G}^+_{-n}|0\rangle$ and $\hat{G}^-_{-n}|0\rangle$ will always differ by a sign. However in case of the $\mathcal{N}=2$ superconformal algebra the norm of these states would be the same. Thus while in the superconformal case it is possible to have states corresponding to the modes $\hat{G}^\pm_{-n}$ which have positive norm this is not possible for the  $\mathcal{W}^{(2)}_3$ algebra. Hence we arrive at the following conclusion:
	\begin{itemize}
		\item\emph{Unless the states $\hat{G}^+_{-n}|0\rangle$ and $\hat{G}^-_{-n}|0\rangle$ are null states there are no unitary representations of the
				 $\mathcal{W}^{(2)}_3$ algebra for our choice of the vacuum given by \eqref{AdSxR:AnnihilationWeight}.}
	\end{itemize}
Looking at \eqref{AdSxR:LevelOnehalfNorm} we see that the only values where $\hat{G}^\pm_{-\frac{3}{2}}|0\rangle$ are null are $\hat{k}=-1$ and $\hat{k}=-\frac{3}{2}$. Choosing one of these two values of $\hat{k}$ does not automatically ensure that the resulting CFT is unitary. One still has to check whether the remaining states in the theory that are not null spoil unitarity or not.\\
In order to simplify the following discussion, it is beneficial to check if the field content of the two CFTs we are looking at is maybe more restricted than one initially thinks. The following discussion applies only to the two values of $\hat{k}$ for which $\hat{G}^\pm_{-n}|0\rangle$ are null states and can thus set
	\begin{equation}
		\hat{G}^\pm_{-n}|0\rangle=0\quad\forall n\in\mathbb{Z}.
	\end{equation} 
This in turn also implies that $[\hat{G}^+_{-n},\hat{G}^-_{-m}]|0\rangle=0$. Thus, also the right hand side of \eqref{AdSxR:W32AlgebraQuantumBershadsky} has to be zero. This leads to the following relation
	\begin{align}\label{AdSxR:LasLinearCombofJ}
		\hat{L}_{-(m+n)}|0\rangle=\frac{1}{\hat{k}+3}\left(\frac{3}{2}(\hat{k}+1)(m-n)J_{-(m+n)}+3\sum_{p>0}^{m+n-1}J_{-p}J_{-(m+n)+p}\right)|0\rangle.
	\end{align}
Thus, we see that for $\hat{k}\in\{-\frac{3}{2},-1\}$ the states $\hat{L}_{-(m+n)}|0\rangle$ are a linear combination of other states which simplifies the theory considerably.\\ 
In order to have a well defined basis of states at level $N$, we employ the following ordering of operators
	\begin{equation}
		J_{-n_1}^{m_1}\ldots J_{-n_{p}}^{m_p}L_{-n_{p+1}}^{m_{p+1}}\ldots L_{-n_N}^{m_N}|0\rangle,
	\end{equation}
with the following restrictions on the indices $m_i$ and $n_i$ 
	\begin{align}
		&m_i\in\mathbb{N},\\
		&n_1,\ldots,n_p\in\mathbb{N}\backslash\{0\},\\
		&n_{p+1},\ldots,n_N\in\mathbb{N}\backslash\{0,1\},\\
		&n_1>\ldots>n_p,\\
		&n_{p+1}>\ldots>n_N,\\
		&\sum_{i=1}^Nm_in_i=N.
	\end{align}
Since $L_{-n_{p+1}}^{m_{p+1}}\ldots L_{-n_N}^{m_N}|0\rangle$ can be rewritten as linear combinations of states of the form $J_{-n_{p+1}}^{m_{p+1}}\ldots J_{-n_N}^{m_N}|0\rangle$ we can express every state at a given level $N$ as
	\begin{equation}\label{AdSxR:LevelNStates}
		J_{-n_1}^{m_1}\ldots J_{-n_N}^{m_N}|0\rangle,
	\end{equation}
where now $n_i\in\mathbb{N}\backslash\{0\}\,\forall i=1,\ldots,N$ and $n_1>\ldots>n_N$.\\
It is also possible to write all states at a given level $N$ as $L_{-1}$ and $J_{-1}$ descendants of level $N-1$ states. For the integer valued levels this can most easily be seen at level 2 and level 1.  At level 1 there is only the state $J_{-1}|0\rangle$ present. Acting with either $L_{-1}$ or $J_{-1}$ on that state one gets the following two states
	\begin{align}
		L_{-1}J_{-1}|0\rangle=&[L_{-1},J_{-1}]|0\rangle=J_{-2}|0\rangle,\\
		J_{-1}J_{-1}|0\rangle=&J_{-1}^2|0\rangle,
	\end{align}
which are all possible states at level 2. One can repeat this process indefinitely and obtain in such a way all possible states at a given level $N$. Please note that with this way of generating states, starting with $M$ states at level $N$ would generate $2M$ states for level $N+1$. Since the number of possible states at level $N$ is given by the number of possible partitions of $N$, this procedure will in general produce "too many" states. This usually happens when a state $|a;N+1\rangle$ at level $N+1$ can be generated by the action of either $L_{-1}$ or $J_{-1}$ on two different states $|b;N\rangle$, $|c;N\rangle$ at level $N$ i.e.
	\begin{align}
		|a;N+1\rangle\propto L_{-1}|b;N\rangle,\\
		|a;N+1\rangle\propto J_{-1}|c;N\rangle.
	\end{align}
Since this will only manifest as additional zero eigenvalues of the Gramian matrix, this would not spoil any unitarity analysis. Thanks to this procedure of generating states it is possible to write \emph{all} states as descendants of the lowest level states, which is very convenient. If for example all states at a given level $N$ are null states, then one can show that all states at levels $M>N$ are also null states. This just follows from the fact that one can write all states at level $N+1$ as descendants of level $N$ states, which are again null states, since descendants of null states are also null states.\\
 The same arguments regarding descendant states also apply to the half integer valued levels. 
\subsubsection{$\hat{k}=-\frac{3}{2}$ and $\hat{c}=0$}
This case is trivial: The only state present in our Hilbert space is the vacuum state itself. This can  easily be seen by looking at \eqref{AdSxR:Level1Norm} and \eqref{AdSxR:LevelOnehalfNorm}, which are null for this value of $\hat{k}$. Thus, all states at integer level $n>1$ and half integer level $m>\frac{3}{2}$ are null, since all states at a given level $m,n$ can be written as descendants of the level 1 and level $\frac{1}{2}$ states. This leads to the conclusion stated at the beginning that the only state present in the theory is the vacuum state. This statement is true in general for the quantum $\mathcal{W}^{(2)}_3$ algebra, if one only has \eqref{AdSxR:W32AlgebraQuantumBershadsky} as a starting point where the only restriction on $\hat{k}$ is that $\hat{k}\neq-3$. However, we obtained this algebra starting from a Chern-Simons action with original level $k$. This gives us another quick argument as to why the resulting CFT is trivial for this value of $\hat{k}$. Looking at \eqref{AdSxR:KGShift} we see that $k=0$ for $\hat{k}=-\frac{3}{2}$. We see that already at the level of the action \eqref{Intro:ICS} the theory is trivial because the action itself is zero for this value of $\hat{k}$.
\subsubsection{$\hat{k}=-1$ and $\hat{c}=1$}
For this value all half integer valued levels contain again only null states. This can again easily be seen by the same argument used for the case $\hat{k}=\frac{3}{2}$. Thus, the only states remaining are the states at integer valued levels since the norm of the level 1 state is positive for this value of $\hat{k}$. In order to check if negative norm states appear at higher levels, we need to calculate the Gramian matrix for any level $N$. Since a general state at a given level $N$ is given by \eqref{AdSxR:LevelNStates}, the entries of the level $N$ Gramian matrix will be given by
	\begin{equation}
		\langle0|J_{\bar{n}_N}^{\bar{m}_N}\ldots J_{\bar{n}_1}^{\bar{m}_1}J_{-n_1}^{m_1}\ldots J_{-n_N}^{m_N}|0\rangle.
	\end{equation}
Using the algebraic relations \eqref{AdSxR:W32AlgebraQuantumBershadsky1} and the property \eqref{AdSxR:AnnihilationWeight} one can immediately see that these entries will be zero unless $\bar{n}_i=n_i\,\forall i=1,\ldots,N$ and $\bar{m}_i=m_i\,\forall i=1,\ldots,N$. Hence we see that the Gramian matrix at level $N$ will always be diagonal and thus the eigenvalues will be equal to the norm of the states present at level $N$. Calculating these norms yield\footnote{The Gramian matrices of the levels 1, $\frac{3}{2}$, 2, 3, 4, 5 as well as a general expression for arbitrary integer valued level $N$ can be found in Appendix \ref{Appendix:Shapovalov}.}
	\begin{equation}
		\langle0|J_{n_N}^{m_N}\ldots J_{n_1}^{m_1}J_{-n_1}^{m_1}\ldots J_{-n_N}^{m_N}|0\rangle=\prod_{i=1}^Nm_i!n_i\left(\frac{2\hat{k}+3}{3}\right)^{m_i}=
		\prod_{i=1}^Nm_i!n_i\left(\frac{1}{3}\right)^{m_i}.
	\end{equation}
Since the eigenvalues of the Gramian matrix are all positive, we see that all states in our theory have positive norm and thus we have a unitary theory for $\hat{k}=-1$ and $\hat{c}=1$.\\
Having found a unitary theory for $\hat{k}=-1$ one can now ask what kind of CFT this is. Looking at \eqref{AdSxR:LasLinearCombofJ} we see that for $\hat{k}=-1$ this expression reduces to a Sugawara construction of the the Virasoro modes $\hat{L}_n$ via the $\mathfrak{u}(1)$ currents $J_n$. i.e.
	\begin{equation}
		\hat{L}_{-n}|0\rangle=\frac{3}{2}\sum_{p>0}^{n-1}J_{-p}J_{-n+p}|0\rangle=\frac{3}{2}\sum_{p\in\mathbb{Z}}:J_{-n-p}J_p:|0\rangle.
	\end{equation} 
Since we obtained $\hat{L}_n$ via a shift given by \eqref{AdSxR:NormOrdShiftVir}, this in turn also means that the unshifted Virasoro modes $L_n$ annihilate the vacuum $\forall n\in\mathbb{Z}$. Thus, the resulting theory looks very similar to a theory based on a $\mathfrak{u}(1)$ current algebra given by
	\begin{equation}\label{AdSxR:JJCommutatorU(1)}
		[J_n,J_m]=n\kappa\delta_{m+n,0}
	\end{equation}
and the following Sugawara construction for the Virasoro modes
	\begin{equation}
		\hat{L}_n=\frac{1}{2\kappa}\sum_{p\in\mathbb{Z}}:J_{-n-p}J_p:.
	\end{equation}
This would yield the following algebra
	\begin{subequations}\label{AdSxR:U1SugawaraConst}
		\begin{align}
			&[J_n,J_m]=n\kappa\delta_{n+m,0},\\
			&[J_n,\hat{L}_m]=nJ_{n+m},\\
			&[\hat{L}_n,\hat{L}_m]=(n-m)\hat{L}_{m+n}+\frac{\hat{c}_\kappa}{12}n(n^2-1)\delta_{n+m,0},
		\end{align}
	\end{subequations}
with $\hat{c}=1$. For $\kappa=\frac{1}{3}$ one would obtain the same field content as for the $\mathcal{W}_3^{(2)}$ algebra for $\hat{k}=-1$. There is, however, a crucial difference between these two theories. While in the case of \eqref{AdSxR:U1SugawaraConst} the parameter $\kappa$ can take arbitrary values\footnote{The only restriction is that $\kappa>0$ for a unitary theory.}, the central charge $\hat{c}_\kappa$ remains unaffected by a change of $\kappa$. Thus, for this theory there is no preferred value of $\kappa$. In the case of the $\mathcal{W}_3^{(2)}$ algebra on the other hand there is a preferred value of $\hat{k}$ and the central charge $\hat{c}$ is also not independent of $\hat{k}$.\\
One way to understand this is to think back to the canonical analysis and the constraints associated with the states  $\hat{G}^\pm_{-n}|0\rangle$. Looking at \eqref{AdSxR:W32AlgebraQuantumBershadsky6}  one can see that for $\hat{k}=-1$ (and $\hat{k}=-\frac{3}{2}$) the central term vanishes. This in turn means that the constraints associated with $\hat{G}^\pm_{-n}|0\rangle$ remain first class even at the boundary and thus there is an additional gauge symmetry present at the boundary that constrains our theory\footnote{In the case of $\hat{k}=-\frac{3}{2}$ all central extensions in the $\mathcal{W}_3^{(2)}$ algebra vanish which is another reason why this theory is trivial. All first class constraints remain first class at the boundary and thus the theory has no degrees of freedom left after fixing the gauge.}. This additional gauge symmetry then restricts the normalization of the $J_n$ modes and forces $\kappa$ to take the value $\frac{1}{3}$. Thus demanding unitarity of the quantum theory leads to an additional symmetry enhancement of the CFT which in turn further constrains the theory.\\ 
In order to give this fixed value of $\kappa$ a physical interpretation one can consider for example a free boson defined on the complex plane described by the action \cite{Blumenhagen}
	\begin{equation}\label{AdSxR:FreeBosonAction}
		S=\frac{1}{4\pi\kappa}\int\extd z\extd \bar{z}\,\partial X\cdot\bar{\partial}X,
	\end{equation}
where $\partial X\cdot\bar{\partial}X$ denotes $\sqrt{|g|}g^{ab}\partial_aX\bar{\partial}_bX$, with $g_{ab}\extd x^a\extd x^b=\frac{\extd z\extd\bar{z}}{z\bar{z}}$. One can then define the following chiral [anti-chiral] fields\footnote{This follows from the equations of motion $\partial\bar{\partial}X(z,\bar{z})=0$.} with conformal weight 1
	\begin{equation}
		j(z)=i\partial X(z,\bar{z}),\quad
		\bar{j}(\bar{z})=i\bar{\partial}X(z,\bar{z}),
	\end{equation}
whose Laurent modes, which are given by
	\begin{equation}
		j(z)=\frac{1}{2\pi}\sum_{n\in\mathbb{Z}}J_nz^{-n-1},\quad\bar{j}(\bar{z})=\frac{1}{2\pi}\sum_{n\in\mathbb{Z}}\bar{J}_n\bar{z}^{-n-1},
	\end{equation}
obey the commutation relations given by \eqref{AdSxR:JJCommutatorU(1)}. Since $\kappa$ appearing in \eqref{AdSxR:JJCommutatorU(1)} is the same as in \eqref{AdSxR:FreeBosonAction} this parameter is essentially the coupling constant of this theory. Thus, demanding unitarity and nontriviality of the CFT based on a $\mathcal{W}_3^{(2)}$ algebra would fix this coupling constant to a certain value and one could interpret the resulting theory as a free boson with a coupling constant fixed by an additional gauge symmetry.

\clearpage

\section{Conclusion}
\begin{figure}[h]
	\centering
		\includegraphics[width=\textwidth]{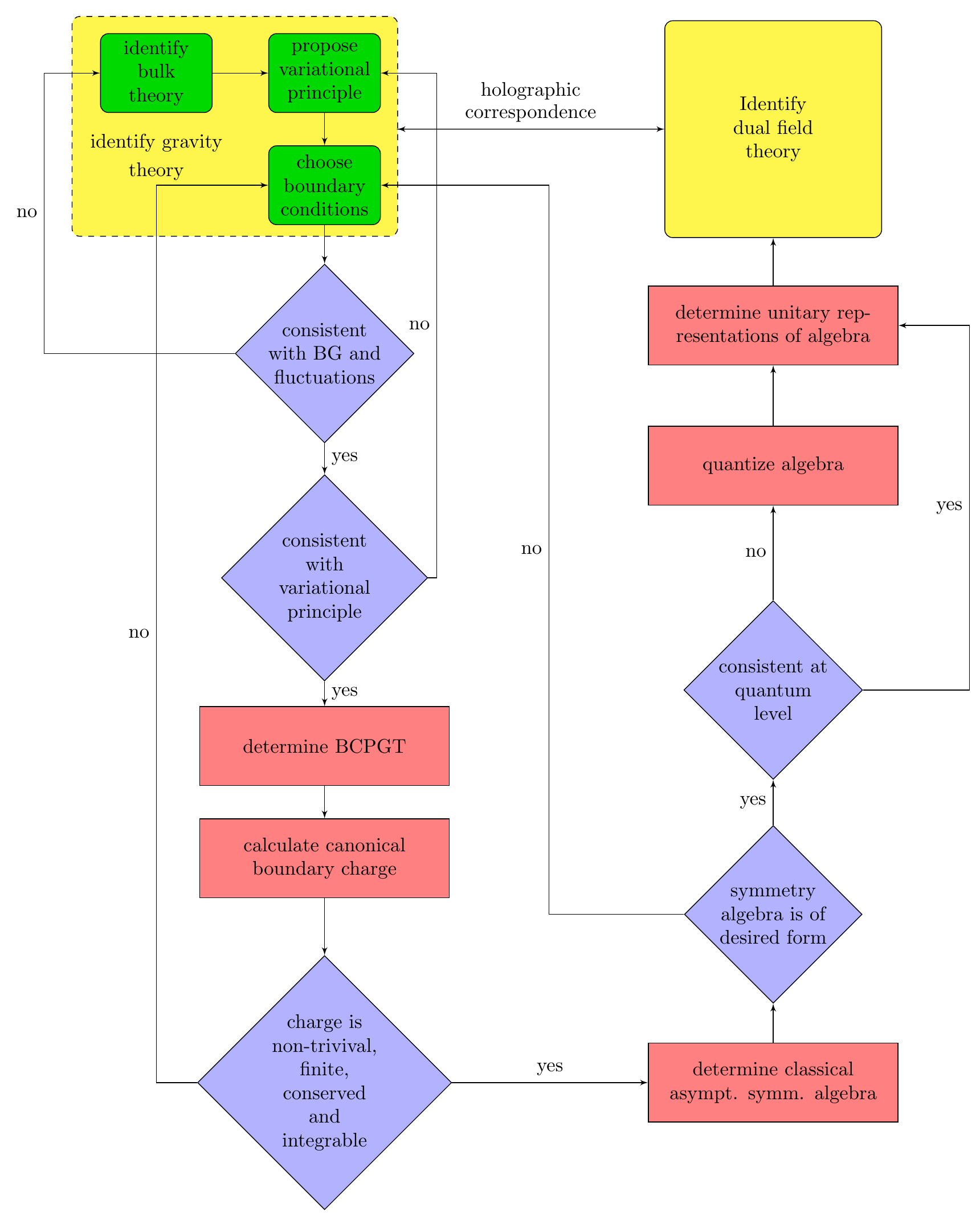}
	\caption{Flowchart depicting the procedure of analyzing higher-spin gravity theories}
	\label{fig:procedure}
\end{figure}

The goal of this thesis was to analyze asymptotical symmetry algebras of (2+1)-dimensional non-AdS higher-spin gravity with a focus on AdS$_2\times\mathbb{R}$ and $\mathbb{H}_2\times\mathbb{R}$. We found a consistent set of boundary conditions that yield finite, integrable, conserved and nontrivial boundary charges. Then we determined the classical symmetry algebra of these boundary charges and found a classical $\mathcal{W}_3^{(2)}\oplus\mathfrak{u}(1)$ symmetry algebra at the boundary.\\ 
Since we were also curious to find out what kind of CFT this symmetry algebra would yield, we tried to obtain the quantized version of the $\mathcal{W}_3^{(2)}$ algebra by satisfying the Jacobi identities. Analyzing the field content of the quantized version of the $\mathcal{W}_3^{(2)}$ algebra and looking for possible unitary representations of the $\mathcal{W}_3^{(2)}$ algebra we found some quite interesting features. For our definition of the vacuum we did find two unitary representations of the $\mathcal{W}_3^{(2)}$ algebra. The reason that we only found two possible unitary representations is that the modes $\hat{G}^\pm_n$ appearing in the $\mathcal{W}_3^{(2)}$ algebra are bosonic and hence obey commutator relations. Because of this the norms of the modes $\hat{G}^+_{-n}$ and $\hat{G}^-_{-n}$ differ by a sign and hence these states have to be null in order to have a chance of obtaining a unitary theory. Thus, by demanding unitarity, ghost states are automatically projected out of the theory.\\
 Taking a closer look at the two unitary representations that we obtained we found that one of these representations is trivial and the other one is very similar to a theory described via $\mathfrak{u}(1)$ currents. However, in the case of the $\mathcal{W}_3^{(2)}$ algebra the resulting theory is more restricted than for the pure $\mathfrak{u}(1)$ case. The reason for this is an additional gauge symmetry enhancement that occurs at the two special values of $\hat{k}$ where the states corresponding to $\hat{G}^\pm_{-n}$ are null. This gauge symmetry enhancement could also explain the value of certain coupling constants such as for example the free boson as we suggested.\\
Having this example of a theory where the coupling constant is fixed by an enhanced gauge symmetry originating from a higher-spin algebra it is tempting to think that this might also work for other theories. One could for example try to do the same analysis for AdS$_2\times\mathbb{R}$ and $\mathbb{H}_2\times\mathbb{R}$ as we did in this thesis, but for spin-4 gravity and the 2-2 embedding of $\mathfrak{sl}(4)$. For the interested reader we provided a suitable basis in appendix \ref{Appendix:2-2}.\\
Another interesting question is related to the two values of $\hat{k}$, for which the resulting CFT is unitary. It could be possible that there is another choice of vacuum for which we have more values of $\hat{k}$ that allow unitary representations. This could for example be realized by different embeddings of the non principal $\mathfrak{sl}(3)$ embeddings in $\mathfrak{sl}(N)$, with $N>3$. One could for example consider the 2-1-1 embedding\footnote{A suitable basis is given in appendix \ref{Appendix:2-1-1}.} of $\mathfrak{sl}(4)$ and again perform the same analysis as in this thesis. At first glance it seems like this analysis is straightforward and should be analog to the spin-3 case that we analyzed in this thesis.  After having determined the quantum asymptotic symmetry algebra, one would then have to check whether or not the $\mathfrak{sl}(4)$ invariant vacuum allows more values of $\hat{k}$, for which the $\mathcal{W}_3^{[2]}$ algebra that should be contained in the resulting asymptotic symmetry algebra is unitary. In principle it could also be possible that there are even less possible values of $\hat{k}$ since there are more Jacobi identities to be fulfilled for the quantum version of the asymptotic symmetry algebra.\\
Since the $\mathcal{W}_3^{[2]}$ algebra is very similar to the $\mathcal{N}=2$ superconformal algebra, it is also interesting to check what happens if we start the canonical analysis with a supersymmetric theory rather than a $\mathfrak{sl}(3)$ invariant theory. One would again expect a correlation between the unshifted Chern-Simons level $k$ and the central charge $c$. Since in the supersymmetric case the modes $\hat{G}^\pm_{n}$ obey anticommutation relations and thus the norms of the states corresponding to $\hat{G}^+_{-n}$ and $\hat{G}^-_{-n}$ have the same sign, there should be a wider range of $k$ that allow for unitary representations.\\
The specific example of AdS$_2\times\mathbb{R}$ [$\mathbb{H}_2\times\mathbb{R}$] and the non-principal embedding of $\mathfrak{sl}(3)$ provided in this thesis can also be used to employ a general procedure in analyzing higher-spin gravity theories formulated via a $SL(N)\times SL(N)$ Chern-Simons formulation \cite{Afshar:2012nk}.\\
One important starting point of this analysis is the correct choice of embedding of $\mathfrak{sl}(2)$ in $\mathfrak{sl}(N)$. If the embedding cannot reproduce the chosen background, then it will be impossible to find boundary conditions that are consistent with the background and the fluctuations. Thus, if a chosen set of boundary conditions is not consistent with the background and the fluctuations, then the reason for this is not necessarily a bad choice of boundary conditions. The inconsistencies could also be the result of a bad choice of embedding or spin-$N$ theory. We will list in the following the basic steps one has to do in order to analyze higher-spin gravity theories. A more detailed version of this procedure is given by figure \ref{fig:procedure} in terms of a flowchart.
	\begin{enumerate}
		\item\label{Conclusion:0} Identify the bulk theory and propose a variational principle.
		\item\label{Conclusion:1} Choose boundary conditions of the connections $A$ and $\bar{A}$ that lead to the desired background (BG) solution and are compatible with a given set of fluctuations of the BG and the variational principle employed.
		\item\label{Conclusion:2} Determine the boundary condition preserving gauge transformations (BCPGT).
		\item\label{Conclusion:3} Calculate the canonical boundary charge.
		\item\label{Conclusion:4} Determine the classical asymptotic symmetry algebra.
		\item\label{Conclusion:5} Quantize the classical asymptotic symmetry algebra if necessary. This can be done for example by introducing normal ordering and imposing the Jacobi identities on the quantum level.
		\item\label{Conclusion:6} Analyze the field content of the resulting CFT.
	\end{enumerate}

\clearpage

\begin{appendix}
\section{Suitable Spin-3 Bases}
For the  $\mathfrak{sl}(2)$ generators we use the following conventions
	\begin{equation}
		[L_n,L_m]=(n-m)L_{n+m}
	\end{equation}
where $L_{\pm1}:=L_{\pm}$. The commutation relations of the remaining generators of the $\mathcal{W}$-Algebras are given by
	\begin{equation}
		[L_n,W_m^{l[a]}]=(nl-m)W_{n+m}^{l[a]}.
	\end{equation}
The index $l$ appearing in $W_m^{l[a]}$ is an $\mathfrak{sl}(2)$ quantum number, while $[a]$ is a color index.
The traces of these generators are given by 
	\begin{equation}
		\textnormal{tr}(W_m^{k[a]}W_n^{l[b]})=(-1)^{l-m}\frac{(l+m)!(l-m)!}{2l!}\delta^{k,l}\delta_{m+n,0}N^{a,b}_l
	\end{equation}
with the normalization
	\begin{equation}
		N^{a,b}_l:=\textnormal{tr}(W_l^{l[a]}W_{-l}^{l[b]}).
	\end{equation}
Whenever singlets fall into an  $\mathfrak{sl}(2)$ on their own their generators will be defined such that they obey 
	\begin{equation}
		[S^{[n]},S^{[m]}]=(n-m)S^{[n+m]}.
	\end{equation}
In addition, we also use the notation $S^{[n]}:=W_0^{0[n]}$. If there is only one singlet present in our representation we just denote it by $S$. Doublets are denoted by $\psi^{[a]}_n:=W^{\frac{1}{2}[a]}_n$.
\subsection{Non-Principal Embedding}\label{Appendix:NonPrincipalEmb}
For the non-principal embedding of $\mathfrak{sl}(2)$ in $\mathfrak{sl}(3)$ in section (\ref{AdSxR}) we used the following set of generators obeying the commutation relations given by \ref{AdSxR:CommRel} .
	\begin{equation}
		L_0=\frac{1}{2}\left(
			\begin{array}{ccc}
				1&0&0\\
				0&0&0\\
				0&0&-1
			\end{array}\right)\quad
		L_+=\left(
			\begin{array}{ccc}
				0&0&0\\
				0&0&0\\
				1&0&0
			\end{array}\right)\quad
		L_-=\left(
			\begin{array}{ccc}
				0&0&-1\\
				0&0&0\\
				0&0&0
			\end{array}\right)
	\end{equation}
Doublets:
	\begin{equation}
		\psi_{\frac{1}{2}}^{+}=\left(
			\begin{array}{ccc}
				0&0&0\\
				-1&0&0\\
				0&0&0
			\end{array}\right)\quad
		\psi_{-\frac{1}{2}}^{+}=\left(
			\begin{array}{ccc}
				0&0&0\\
				0&0&1\\
				0&0&0	
			\end{array}\right)	
	\end{equation}
	\begin{equation}
		\psi_{\frac{1}{2}}^{-}=\left(
			\begin{array}{ccc}
				0&0&0\\
				0&0&0\\
				0&1&0
			\end{array}\right)\quad
		\psi_{-\frac{1}{2}}^{-}=\left(
			\begin{array}{ccc}
				0&1&0\\
				0&0&0\\
				0&0&0	
			\end{array}\right)	
	\end{equation}
Singlet:
	\begin{equation}
		S=\frac{1}{3}\left(
			\begin{array}{ccc}
				-1&0&0\\
				0&2&0\\
				0&0&-1
			\end{array}\right)
	\end{equation}
Killing form:
	\begin{equation}
		g_{ab}=\left(
			\begin{array}{cccccccc}
				0&0&-1&0&0&0&0&0\\
				0&\frac{1}{2}&0&0&0&0&0&0\\
				-1&0&0&0&0&0&0&0\\
				0&0&0&0&0&0&-1&0\\
				0&0&0&0&0&1&0&0\\
				0&0&0&0&1&0&0&0\\
				0&0&0&-1&0&0&0&0\\
				0&0&0&0&0&0&0&\frac{2}{3}
			\end{array}\right),
	\end{equation}
with the generators are ordered as $L_1,L_0,L_{-1},\psi_{\frac{1}{2}}^{+},\psi_{-\frac{1}{2}}^{+},\psi_{\frac{1}{2}}^{-},\psi_{-\frac{1}{2}}^{-},S$.
\section{Gramian Matrices for $\hat{k}\in\{-\frac{3}{2},-1\}$}\label{Appendix:Shapovalov}
A general expression for calculating all coefficients for a Gramian matrix at integer valued level $N$ of the $\mathcal{W}^{(2)}_{3}$ algebra is given by
	\begin{equation}
		\langle0|J_{\bar{n}_N}^{\bar{m}_N}\ldots J_{\bar{n}_1}^{\bar{m}_1}J_{-n_1}^{m_1}\ldots J_{-n_N}^{m_N}|0\rangle=
		\prod_{i=1}^Nm_i!n_i\left(\frac{2\hat{k}+3}{3}\right)^{m_i}\delta_{m_i,\bar{m}_i}\delta_{n_i,\bar{n}_i}.
	\end{equation}
Level 2:
	\begin{equation}
		K^{(2)}=\left(
		\begin{array}{cc}
			2\frac{2\hat{k}+3}{3}&0\\
			0&2\left(\frac{2\hat{k}+3}{3}\right)^2			
		\end{array}\right),
	\end{equation}
with the basis vectors arranged as $J_{-2}|0\rangle$, $J_{-1}^2|0\rangle$. \\\\
Level $\frac{5}{2}$:
	\begin{equation}
		K^{(\frac{5}{2})}=(\hat{k}+1)(2\hat{k}+3)\left(
		\begin{array}{cccc}
			-3&0&2&0\\
			0&3&0&2\\
			2&0&-\frac{2}{3}(\hat{k}+3)&0\\
			0&2&0&\frac{2}{3}(\hat{k}+3)
		\end{array}\right),
	\end{equation} 
with the basis vectors arranged as $G^+_{-\frac{5}{2}}|0\rangle$, $G^-_{-\frac{5}{2}}|0\rangle$,  $G^+_{-\frac{3}{2}}J_{-1}|0\rangle$ and  $G^-_{-\frac{3}{2}}J_{-1}|0\rangle$.\\\\
Level 3:
	\begin{equation}
		K^{(3)}=\left(
		\begin{array}{ccc}
			2\hat{k}+3&0&0\\
			0&2\left(\frac{2\hat{k}+3}{3}\right)^2&0\\
			0&0&6\left(\frac{2\hat{k}+3}{3}\right)^3
		\end{array}\right),
	\end{equation}
with the basis vectors arranged as $J_{-3}|0\rangle$, $J_{-2}J_{-1}|0\rangle$, $J_{-1}^3|0\rangle$.\\\\
Level 4:
	\begin{equation}
		K^{(4)}=\left(
		\begin{array}{ccccc}
			4\frac{2\hat{k}+3}{3}&0&0&0&0\\	
			0&3\left(\frac{2\hat{k}+3}{3}\right)^2&0&0&0\\
			0&0&8\left(\frac{2\hat{k}+3}{3}\right)^2&0&0\\
			0&0&0&4\left(\frac{2\hat{k}+3}{3}\right)^3&0\\
			0&0&0&0&24\left(\frac{2\hat{k}+3}{3}\right)^4
		\end{array}\right),
	\end{equation}
with the basis vectors arranged as $J_{-4}|0\rangle$, $J_{-3}J_{-1}|0\rangle$, $J^2_{-2}|0\rangle$, $J_{-2}J_{-1}^2|0\rangle$,  $J_{-1}^4|0\rangle$.\\\\
Level 5:
	\begin{equation}
		K^{(5)}=\left(
		\begin{array}{ccccccc}
			5\frac{2\hat{k}+3}{3}&0&0&0&0&0&0\\	
			0&4\left(\frac{2\hat{k}+3}{3}\right)^2&0&0&0&0&0\\
			0&0&6\left(\frac{2\hat{k}+3}{3}\right)^2&0&0&0&0\\
			0&0&0&6\left(\frac{2\hat{k}+3}{3}\right)^3&0&0&0\\
			0&0&0&0&8\left(\frac{2\hat{k}+3}{3}\right)^3&0&0\\
			0&0&0&0&0&12\left(\frac{2\hat{k}+3}{3}\right)^4&0\\
			0&0&0&0&0&0&120\left(\frac{2\hat{k}+3}{3}\right)^5
		\end{array}\right),
	\end{equation}
with the basis vectors arranged as $J_{-5}|0\rangle$, $J_{-4}J_{-1}|0\rangle$, $J_{-3}J_{-2}|0\rangle$, $J_{-3}J_{-1}^2|0\rangle$, $J_{-2}^2J_{-1}|0\rangle$, $J_{-2}J_{-1}^3|0\rangle$, $J_{-1}^5|0\rangle$.
\section{Suitable Spin-4 Bases}
In this section we present two different embeddings of $\mathfrak{sl}(2)$ in $\mathfrak{sl}(4)$. We use the same notation as in \cite{Gary:2012aa} where one my also find further embeddings for $\mathfrak{sl}(4)$.
\subsection{2-2 Embedding}\label{Appendix:2-2}
$\mathfrak{sl}(2)$ generators:
	\begin{equation}
		L_0=\frac{1}{2}\left(
			\begin{array}{cccc}
				1&0&0&0\\
				0&1&0&0\\
				0&0&-1&0\\
				0&0&0&-1
			\end{array}\right)\quad
		L_+=\left(
			\begin{array}{cccc}
				0&0&0&0\\
				0&0&0&0\\
				1&0&0&0\\
				0&1&0&0
			\end{array}\right)\quad
		L_-=\left(
			\begin{array}{cccc}
				0&0&-1&0\\
				0&0&0&-1\\
				0&0&0&0\\
				0&0&0&0
			\end{array}\right)
	\end{equation}
Other triplets:
	\begin{subequations}
		\begin{align}
		T_0^{[1]}=&\frac{1}{2}\left(
			\begin{array}{cccc}
				1&0&0&0\\
				0&0&0&0\\
				0&0&-1&0\\
				0&0&0&0
			\end{array}\right)\quad
		T_+^{[1]}=&\left(
			\begin{array}{cccc}
				0&0&0&0\\
				0&0&0&0\\
				1&0&0&0\\
				0&0&0&0
			\end{array}\right)\quad
		T_-^{[1]}=&\left(
			\begin{array}{cccc}
				0&0&-1&0\\
				0&0&0&0\\
				0&0&0&0\\
				0&0&0&0
			\end{array}\right)\\
		T_0^{[2]}=&\frac{1}{2}\left(
			\begin{array}{cccc}
				1&1&0&0\\
				0&0&0&0\\
				0&0&-1&-1\\
				0&0&0&0
			\end{array}\right)\quad
		T_+^{[2]}=&\left(
			\begin{array}{cccc}
				0&0&0&0\\
				0&0&0&0\\
				1&1&0&0\\
				0&0&0&0
			\end{array}\right)\quad
		T_-^{[2]}=&\left(
			\begin{array}{cccc}
				0&0&-1&-1\\
				0&0&0&0\\
				0&0&0&0\\
				0&0&0&0
			\end{array}\right)\\
		T_0^{[3]}=&\frac{1}{2}\left(
			\begin{array}{cccc}
				0&0&0&0\\
				1&1&0&0\\
				0&0&0&0\\
				0&0&-1&-1
			\end{array}\right)\quad
		T_+^{[3]}=&\left(
			\begin{array}{cccc}
				0&0&0&0\\
				0&0&0&0\\
				0&0&0&0\\
				1&1&0&0
			\end{array}\right)\quad
		T_-^{[3]}=&\left(
			\begin{array}{cccc}
				0&0&0&0\\
				0&0&-1&-1\\
				0&0&0&0\\
				0&0&0&0
			\end{array}\right)			
		\end{align}
	\end{subequations}
Singlets:
	\begin{equation}
		S^{[0]}=\frac{1}{2}\left(
			\begin{array}{cccc}
				1&0&0&0\\
				0&-1&0&0\\
				0&0&1&0\\
				0&0&0&-1
			\end{array}\right)\quad
		S^{[+]}=\left(
			\begin{array}{cccc}
				0&0&0&0\\
				1&0&0&0\\
				0&0&0&0\\
				0&0&1&0
			\end{array}\right)\quad
		S^{[-]}=\left(
			\begin{array}{cccc}
				0&-1&0&0\\
				0&0&0&0\\
				0&0&0&-1\\
				0&0&0&0
			\end{array}\right)
	\end{equation}
\subsection{2-1-1 Embedding}\label{Appendix:2-1-1}
$\mathfrak{sl}(2)$ generators:
	\begin{equation}
		L_0=\frac{1}{2}\left(
			\begin{array}{cccc}
				1&0&0&0\\
				0&0&0&0\\
				0&0&0&0\\
				0&0&0&-1
			\end{array}\right)\quad
		L_+=\left(
			\begin{array}{cccc}
				0&0&0&0\\
				0&0&0&0\\
				0&0&0&0\\
				1&0&0&0
			\end{array}\right)\quad
		L_-=\left(
			\begin{array}{cccc}
				0&0&0&-1\\
				0&0&0&0\\
				0&0&0&0\\
				0&0&0&0
			\end{array}\right)
	\end{equation}
Doublets:
		\begin{align}
		G_+^{[1]}=&\left(
			\begin{array}{cccc}
				0&0&0&0\\
				0&0&0&0\\
				1&0&0&0\\
				0&0&0&0
			\end{array}\right)\quad
		G_+^{[2]}=&\left(
			\begin{array}{cccc}
				0&0&0&0\\
				0&0&0&0\\
				0&0&0&0\\
				0&0&1&0
			\end{array}\right)\quad
		G_+^{[3]}=&\left(
			\begin{array}{cccc}
				0&0&0&0\\
				1&0&0&0\\
				0&0&0&0\\
				0&0&0&0
			\end{array}\right)\quad
		G_+^{[4]}=&\left(
			\begin{array}{cccc}
				0&0&0&0\\
				0&0&0&0\\
				0&0&0&0\\
				0&-1&0&0
			\end{array}\right)\nonumber\\
		G_-^{[1]}=&\left(
			\begin{array}{cccc}
				0&0&0&0\\
				0&0&0&0\\
				0&0&0&-1\\
				0&0&0&0
			\end{array}\right)\quad
		G_-^{[2]}=&\left(
			\begin{array}{cccc}
				0&0&1&0\\
				0&0&0&0\\
				0&0&0&0\\
				0&0&0&0
			\end{array}\right)\quad
		G_-^{[3]}=&\left(
			\begin{array}{cccc}
				0&0&0&0\\
				0&0&0&-1\\
				0&0&0&0\\
				0&0&0&0
			\end{array}\right)
		G_-^{[4]}=&\left(
			\begin{array}{cccc}
				0&-1&0&0\\
				0&0&0&0\\
				0&0&0&0\\
				0&0&0&0
			\end{array}\right)	
		\end{align}
Singlets:
	\begin{subequations}
		\begin{align}
		S^{[0]}=\frac{1}{2}\left(
			\begin{array}{cccc}
				0&0&0&0\\
				0&1&0&0\\
				0&0&-1&0\\
				0&0&0&0
			\end{array}\right)\quad
		S^{[+]}=&\left(
			\begin{array}{cccc}
				0&0&0&0\\
				0&0&0&0\\
				0&1&0&0\\
				0&0&0&0
			\end{array}\right)\quad
		S^{[-]}=\left(
			\begin{array}{cccc}
				0&0&0&0\\
				0&0&-1&0\\
				0&0&0&0\\
				0&0&0&0
			\end{array}\right)\\
		S=&\left(
			\begin{array}{cccc}
				1&0&0&0\\
				0&-1&0&0\\
				0&0&-1&0\\
				0&0&0&1
			\end{array}\right)
		\end{align}
	\end{subequations}
\end{appendix}

\clearpage

\bibliographystyle{fullsort}
\bibliography{review}

\end{document}